%% file: TAC-RelObs.tex
\documentclass[journal,11pt,draftclsnofoot,onecolumn]{IEEEtran}

\usepackage{graphicx}

\usepackage[sort,compress]{cite}

\usepackage{latexsym}
\usepackage{psfrag}
\usepackage{amsthm,amssymb,amscd,amssymb,amsmath,mathrsfs}
\usepackage[noend]{algpseudocode}
\usepackage{multirow}
\usepackage{color}

\theoremstyle{remark}

\newtheorem{lem}{Lemma}
\newtheorem{thm}{Theorem}
\newtheorem{defn}{Definition}

\newtheorem{prop}{Proposition}

\newtheorem{exmp}{Example}

\begin{document}

\title{Relative Observability of Discrete-Event Systems and its Supremal Sublanguages}

\author{Kai Cai, Renyuan Zhang, and W.M. Wonham 
\thanks{K. Cai is with Urban Research Plaza, Osaka City University, Japan.
R. Zhang is with Department of Traffic and Control Engineering,
Northwestern Polytechnical University, China. W.M. Wonham is with
Department of Electrical and Computer Engineering, University of Toronto, Canada. 
This work was supported in part by the Natural Sciences and
Engineering Research Council, Canada, Grant no. 7399, and Program to
Disseminate Tenure Tracking System, MEXT, Japan.}

}

\maketitle

\begin{abstract}
We identify a new observability concept, called \emph{relative
observability}, in supervisory control of discrete-event systems
under partial observation.  A fixed, ambient language is given,
relative to which observability is tested.  Relative observability
is stronger than observability, but enjoys the important property
that it is preserved under set union; hence there exists the
supremal relatively observable sublanguage of a given language.
Relative observability is weaker than normality, and thus yields,
when combined with controllability, a generally larger controlled
behavior; in particular, no constraint is imposed that only
observable controllable events may be disabled. We design new algorithms
which compute the supremal relatively observable (and controllable)
sublanguage of a given language, which is generally larger than the
normal counterparts. We demonstrate the new observability concept
and algorithms with a Guideway and an AGV example.
\end{abstract}


\input{sec_intro}

\input{sec_relobs}


\input{sec_relobs-gen}


\input{sec_relobs-con}

\input{sec_examples}


\input{sec_concl}

\bibliographystyle{IEEEtran}
\bibliography{observable,dist_cont,decen_sup,fundam,DistributedControl}




\end{document}

%% file: sec_intro.tex
\section{Introduction} \label{sec_intro}

\begin{figure}[!t]
  \centering
  \includegraphics[width=0.48\textwidth]{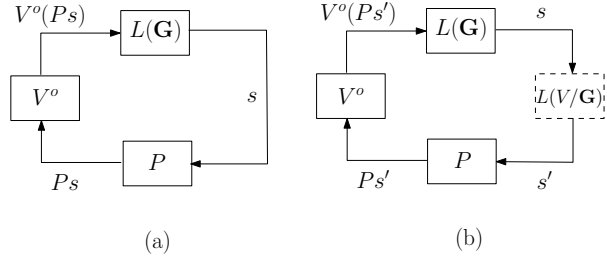}
  \caption{Supervisory control under partial observation. $L({\bf G})$ is the closed behavior of
  the plant, $P$ a natural projection modeling the observation channel, $V^o$ the supervisor under partial observation.
  In (b), $L(V/{\bf G})$ is the closed-loop controlled behavior with full observation.}
  \label{fig:partial_obs}
\end{figure}

In supervisory control of discrete-event systems, partial
observation arises when the supervisor does not observe all events
generated by the plant \cite{SCDES,CaLa:07}. This situation is
depicted in Fig.~\ref{fig:partial_obs}(a), where ${\bf G}$ is the
plant with closed behavior $L({\bf G})$ and marked behavior
$L_m({\bf G})$, $P$ is a natural projection that nulls unobservable
events, and $V^o$ is the supervisor under partial observation. The
fundamental \emph{observability} concept is identified in
\cite{LinWon:88a,Cieslak:88}: observability and controllability of a
language $K \subseteq L_m({\bf G})$ is necessary and sufficient for
the existence of a \emph{nonblocking} supervisor $V^o$ synthesizing
$K$. The observability property is not, however, preserved under set
union, and hence there generally does not exist the supremal
observable and controllable sublanguage of a given language.

The normality concept studied in \cite{LinWon:88a,Cieslak:88} is
stronger than observability but algebraically well-behaved: there
always exists the supremal normal and controllable sublanguage of a
given language.  The supremal sublanguage may be computed by methods
in \cite{ChoMarcus:89,BrandtLin:90}; also see a coalgebra-based
method in \cite{KomSch:05}. Normality, however, imposes the
constraint that controllable events cannot be disabled unless they
are observable \cite[Section~6.5]{SCDES}. This constraint might
result in overly conservative controlled behavior.

To fill the gap between observability and normality, in this paper
we identify a new concept called \emph{relative observability}. For
a language $K \subseteq L_m({\bf G})$, we fix an \emph{ambient}
language $\overline{C}$ such that $\overline{K} \subseteq
\overline{C} \subseteq L({\bf G})$ (here $\overline{\ \cdot\ }$
denotes \emph{prefix closure}, defined in Section~\ref{sec_relobs}).
It is relative to the ambient language $\overline{C}$ that
observability of $K$ is tested. We prove that relative observability
is stronger than the observability in \cite{LinWon:88a,Cieslak:88}
(strings in $\overline{C}-\overline{K}$, if any, need to be tested),
weaker than normality (unobservable controllable events may be
disabled), and preserved under set union. Hence, there exists the
supremal relatively observable and controllable sublanguage of a
given language, which is generally larger than the supremal normal
counterpart, and may be synthesized by a nonblocking supervisor.
This result is useful in practical situations where there may be not
enough sensors available for all controllable events, or it might be
too costly to have all; the result may also help deal with the
practical issue of sensor failures.

We then design new algorithms to compute the supremal sublanguages, capable of keeping
track of the ambient language. These results are
demonstrated with a Guideway and an AGV example in
Section~\ref{sec_examples}, providing quantitative evidence of improvements
by relative observability as compared to normality.
Note that in the special case $\overline{C} = \overline{K}$,
relative observability coincides with observability for the given
$K$. The difference, however, is that when a family of languages is
considered, the ambient $\overline{C}$ in relative observability is
held fixed. It is this feature that renders relative observability
algebraically well-behaved.

Another special case is when the ambient $\overline{C} = L({\bf
G})$.  As suggested by Fig.~\ref{fig:partial_obs}(a), $L({\bf G})$
is a natural choice for the ambient language because strings in
$L({\bf G})$ are observed through the channel $P$. When control is
in place, a more reasonable choice for the ambient $\overline{C}$ is
$L(V/{\bf G})$, the optimal nonblocking controlled behavior under
full observation, since any string in $L({\bf G})-L(V/{\bf G})$ is
effectively prohibited by control; see
Fig.~\ref{fig:partial_obs}(b). With $\overline{C}=L(V/{\bf G})$, the
supremal relatively observable and controllable sublanguage is
generally larger than the supremal normal counterpart; this is
illustrated by empirical studies on a Guideway and an AGV example in
Section~\ref{sec_examples}.

In \cite{TakaiUshio:03}, Takai and Ushio reported an observability
property, formulated in a state-based form, which is preserved under
a union operation of ``strict subautomata''. This operation does not
correspond to language union. It was shown that the (marked)
language of ``the supremal subautomaton'' with the proposed
observability is generally larger than the supremal normal
counterpart.  As will be illustrated by examples, their
observability property and our relative observability do not
generally imply each other. In the Guideway example in
Subsection~\ref{sec_guideway}, we present a case where our algorithm
computes a strictly larger controlled behavior.

We note that, for prefix-closed languages, several procedures are
developed to compute a maximal observable and controllable
sublanguage, e.g.
\cite{ChoMarcus:MST89,FaYaZh:93,HeyLin:94,HadLafLin:96,Ushio:99}.
Those procedures are not, however, applicable to non-closed
languages, because the resulting supervisor may be blocking. In
addition, the observability concept has been extended to
coobservability in decentralized supervisory control (e.g.
\cite{RudWon:92,YooLaf:02}), state-based observability (e.g.
\cite{LiWon:88,KumGarMar:93}), timed observability in real-time
discrete-event systems (e.g. \cite{LinWon:95,TakaiUshio:06}), and
optimal supervisory control with costs \cite{MarBoiLaf:01}.
Observability and normality have also been used in modular,
decentralized, and coordination control architectures (e.g.
\cite{LinWon:90,SuSchu:10,KomMasSch:11}). In the present paper, we
focus on centralized, monolithic supervision for untimed systems in
the Ramadge-Wonham language framework \cite{SCDES,WonRam:87}, and
leave those extensions of relative observability for future
research.

The rest of this paper is organized as follows.
Section~\ref{sec_relobs} introduces the relative observability
concept, and establishes its properties.
Section~\ref{sec_relobs-gen} presents an algorithm to compute the
supremal relatively observable sublanguage of a given language, and
Section~\ref{sec_relobs-con} combines relative observability and
controllability to generate controlled behavior generally larger
than the normality counterpart. Section~\ref{sec_examples}
demonstrates the results with a Guideway and an AGV example.
Finally Section~\ref{sec_concl} states our conclusions.


%% file: sec_relobs.tex
\section{Relative Observability} \label{sec_relobs}

The plant to be controlled is modeled by a generator
\begin{align} \label{eq:generator}
\textbf{G} = (Q, \Sigma, \delta, q_0, Q_m)
\end{align}
where $Q$ is the finite state set; $q_0 \in Q$ is the initial state;
$Q_m \subseteq Q$ is the subset of marker states; $\Sigma$ is the
finite event set; $\delta: Q \times \Sigma \rightarrow Q$ is the
(partial) state transition function. In the usual way, $\delta$ is
extended to $\delta: Q \times \Sigma^* \rightarrow Q$, and we write
$\delta(q,s)!$ to mean that $\delta(q,s)$ is defined. The
\emph{closed behavior} of $\textbf{G}$ is the language
\begin{align} \label{eq:closedlang}
L(\textbf{G}) := \{s \in \Sigma^* | \delta(q_0,s)!\} \subseteq
\Sigma^*;
\end{align}
the \emph{marked behavior} is
\begin{align} \label{eq:markedlang}
L_m(\textbf{G}) := \{s \in L(\textbf{G}) | \delta(q_0,s) \in Q_m\}
\subseteq L(\textbf{G}).
\end{align}
A string $s_1$ is a \emph{prefix} of a string $s$, written $s_1 \leq
s$, if there exists $s_2$ such that $s_1 s_2 = s$. The
\emph{(prefix) closure} of $L_m(\textbf{G})$ is
$\overline{L_m(\textbf{G})} := \{ s_1 \in \Sigma^* \ |\ (\exists s
\in L_m(\textbf{G})) s_1 \leq s \}$. In this paper we assume
$\overline{L_m(\textbf{G})} = L(\textbf{G})$; namely $\textbf{G}$ is
\emph{nonblocking}.

For partial observation, let the event set $\Sigma$ be partitioned
into $\Sigma_o$, the observable event subset, and $\Sigma_{uo}$, the
unobservable subset (i.e. $\Sigma = \Sigma_o \dot{\cup}
\Sigma_{uo}$). Bring in the \emph{natural projection} $P : \Sigma^*
\rightarrow \Sigma_o^*$ defined according to
\begin{equation} \label{eq:natpro}
\begin{split}
P(\epsilon) &= \epsilon, \ \ \epsilon \mbox{ is the empty string;} \\
P(\sigma) &= \left\{
  \begin{array}{ll}
    \epsilon, & \hbox{if $\sigma \notin \Sigma_o$,} \\
    \sigma, & \hbox{if $\sigma \in \Sigma_o$;}
  \end{array}
\right.\\
P(s\sigma) &= P(s)P(\sigma),\ \ s \in \Sigma^*, \sigma \in \Sigma.
\end{split}
\end{equation}
In the usual way, $P$ is extended to $P : Pwr(\Sigma^*) \rightarrow
Pwr(\Sigma^*_o)$, where $Pwr(\cdot)$ denotes powerset. Write $P^{-1}
: Pwr(\Sigma^*_o) \rightarrow Pwr(\Sigma^*)$ for the
\emph{inverse-image function} of $P$. Given two languages $L_i
\subseteq \Sigma^*_i$, $i=1,2$, their \emph{synchronous product} is
$L_1 || L_2 := P_1^{-1}L_1 \cap P_2^{-1}L_2 \subseteq (\Sigma_1 \cup
\Sigma_2)^*$, where $P_i : (\Sigma_1 \cup \Sigma_2)^* \rightarrow
\Sigma^*_i$.

\begin{figure}[!t]
  \centering
  \includegraphics[width=0.45\textwidth]{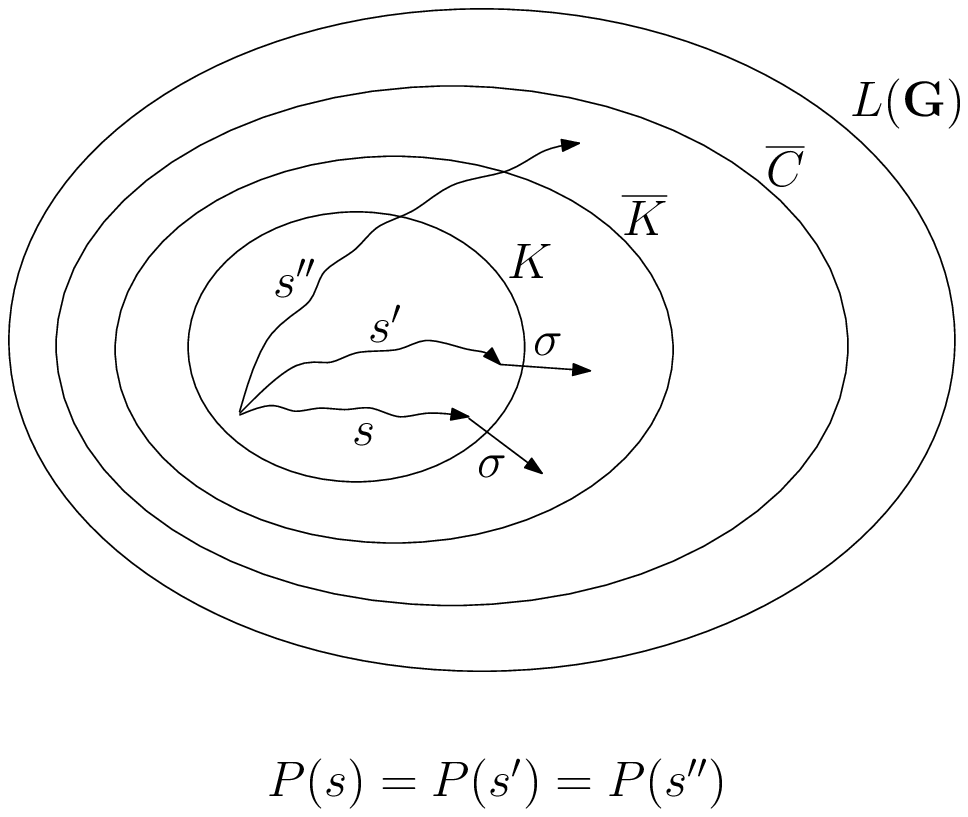}
  \caption{Verification of relative observability of $K$ requires
  checking all three lookalike strings $s, s', s''$ in the ambient
  language $\overline{C}$, while verification of observability of $K$
  requires checking only $s, s'$ in $\overline{K}$.  For $K$ to be
  $\overline{C}$-observable, condition~(\ref{eq:c-obs1}) requires
  $s''\sigma \notin L(\textbf{G})$, and condition~(\ref{eq:c-obs2})
  requires $s'' \notin L_m(\textbf{G})$.}
  \label{fig:c-obs_graph}
\end{figure}

Observability of a language is a familiar concept
\cite{LinWon:88a,Cieslak:88}. Now fixing a sublanguage $C \subseteq
L_m(\textbf{G})$, we introduce \emph{relative observability} which
sets $\overline{C} \subseteq L(\textbf{G})$ to be the \emph{ambient
language} in which observability is tested.
\begin{defn} \label{defn:c-obs}
Let $K \subseteq C \subseteq L_m(\textbf{G})$.  We say $K$ is
\emph{relatively observable} with respect to $\overline{C}$,
$\textbf{G}$, and $P$, or simply $\overline{C}$\emph{-observable},
if for every pair of strings $s, s' \in \Sigma^*$ that are lookalike
under $P$, i.e. $P(s) = P(s')$, the following two conditions hold:
{\small
\begin{align}\label{eq:c-obs1} &(i)\ (\forall \sigma \in \Sigma)\
s\sigma \in \overline{K}, s' \in
\overline{C}, s'\sigma \in L(\textbf{G}) \Rightarrow s'\sigma \in \overline{K} \\
&(ii)\ s \in K, s' \in \overline{C} \cap L_m(\textbf{G}) \Rightarrow
s' \in K \label{eq:c-obs2}
\end{align}}
\end{defn}
Note that a pair of lookalike strings $(s,s')$ trivially satisfies
(\ref{eq:c-obs1}) and ({\ref{eq:c-obs2}}) if either $s$ or $s'$ does
not belong to the ambient $\overline{C}$.  For a lookalike pair
$(s,s')$ both in $\overline{C}$, relative observability requires
that (i) $s$ and $s'$ have identical one-step
continuations,\footnote{Here we consider all one-step transitions
$\sigma \in \Sigma$ because we wish to separate the issue of
observation from that of control.  If and when control is present,
as we will discuss below in Section~\ref{sec_relobs-con}, then we
need to consider only controllable transitions in (\ref{eq:c-obs1})
inasmuch as the controllability requirement prevents uncontrollable
events from violating (\ref{eq:c-obs1}).} if allowed in
$L(\textbf{G})$, with respect to membership in $\overline{K}$; and
(ii) if each string is in $L_m(\textbf{G})$ and one actually belongs
to $K$, then so does the other. A graphical explanation of the
concept is given in Fig.~\ref{fig:c-obs_graph}.

If $\overline{C_1} \subseteq \overline{C_2} \subseteq L({\bf G})$
are two ambient languages, it follows easily from
Definition~\ref{defn:c-obs} that $\overline{C_2}$-observability
implies $\overline{C_1}$-observability. Namely, the smaller the
ambient language, the weaker the relative observability. In the
special case where the ambient $\overline{C}=\overline{K}$,
Definition~\ref{defn:c-obs} becomes the standard observability
\cite{LinWon:88a,Cieslak:88} for the given $K$.  This immediately
implies

\begin{prop} \label{prop:c-obs_obs}
If $K \subseteq  C$ is $\overline{C}$-observable, then $K$ is also
observable.
\end{prop}

\begin{figure}[!t]
  \centering
  \includegraphics[width=0.45\textwidth]{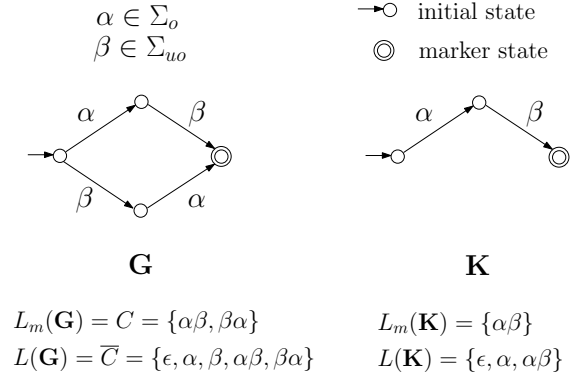}
  \caption{$L_m(\textbf{K})$ is observable but not relatively observable. In $L(\textbf{K})$ the only lookalike string
  pair is $(\alpha, \alpha \beta)$; it is easily verified that $L_m(\textbf{K})$ is observable. To see that $L_m(\textbf{K})$ is
  not $\overline{C}$-observable, let $s=\epsilon$ and $s'=\beta$ ($\notin L(\textbf{K})$). We have $s\alpha \in L(\textbf{K})$,
  $s'\alpha \in \overline{C}=L(\textbf{G})$, but $s'\alpha \notin L(\textbf{K})$. This violates (\ref{eq:c-obs1}). Also
  consider $s=\alpha \beta$ and $s'=\beta \alpha$ ($\notin L_m(\textbf{K})$). We have $s \in L_m(\textbf{K})$,
  $s' \in \overline{C} \cap L_m(\textbf{G})$, but $s' \notin L_m(\textbf{K})$. This violates (\ref{eq:c-obs2}). }
  \label{fig:c-obs_obs}
\end{figure}

The reverse statement need not be true.  An example is provided in
Fig.~\ref{fig:c-obs_obs}, which displays an observable language that
is not relatively observable.

An important way in which relative observability differs from
observability is the exploitation of a fixed ambient $\overline{C}
\subseteq L(\textbf{G})$.  Let $K_i \subseteq C$, $i=1,2$. For
(standard) observability of each $K_i$, one checks lookalike string
pairs only in $\overline{K_i}$, ignoring all candidates permitted by
the other language. Observability of $K_i$ is in this sense
`myopic', and consequently, both $K_i$ being observable need not
imply that their union $K_1 \cup K_2$ is observable.  The fixed
ambient language $\overline{C}$, by contrast, provides a `global
reference': no matter which $K_i$ one checks for relative
observability, all lookalike string pairs in $\overline{C}$ must be
considered. This more stringent requirement renders relative
observability algebraically well-behaved, as we will see in
Subsection~\ref{subsec:relobs-sup}. Before that, we first show the
relation between relative observability and \emph{normality}
\cite{LinWon:88a,Cieslak:88}.

\subsection{Relative observability is weaker than normality} \label{subsec:relobs-normal}

In this subsection, we show that relative observability is weaker
than \emph{normality}, a property that is also preserved by set
unions \cite{LinWon:88a,Cieslak:88}. A sublanguage $K \subseteq C$
is $(L_m(\textbf{G}),P)$\emph{-normal} if
\begin{align} \label{eq:normality}
K = P^{-1} P K \cap L_m(\textbf{G}).
\end{align}
If, in addition, $\overline{K}$ is $(L(\textbf{G}),P)$-normal, then
no string in $\overline{K}$ may exit $\overline{K}$ via an
unobservable transition \cite[Section~6.5]{SCDES}. This means, when
control is present, that one cannot disable any unobservable,
controllable events. Relative observability, by contrast, does not
impose this restriction, i.e. one may exercise control over
unobservable events.

\begin{prop} \label{prop:c-obs_norm}
If $K \subseteq C$ is $(L_m(\textbf{G}),P)$-normal and
$\overline{K}$ is $(L(\textbf{G}),P)$-normal, then $K$ is
$\overline{C}$-observable.
\end{prop}

\emph{Proof.} Let $s,s' \in \Sigma^*$ and $Ps = Ps'$.  We must show
that both (\ref{eq:c-obs1}) and (\ref{eq:c-obs2}) hold for $K$.

For (\ref{eq:c-obs1}), let $\sigma \in \Sigma$, $s \sigma \in
\overline{K}$, $s' \in \overline{C}$, and $s' \sigma \in
L(\textbf{G})$; it will be shown that $s' \sigma \in \overline{K}$.
From $s \sigma \in \overline{K}$ we have
\begin{align*}
P(s \sigma) \in P\overline{K} &\Rightarrow P(s) P(\sigma) \in
P\overline{K} \\
&\Rightarrow P(s') P(\sigma) \in
P\overline{K} \\
&\Rightarrow P(s' \sigma) \in
P\overline{K} \\
&\Rightarrow s'\sigma \in P^{-1}P\overline{K}
\end{align*}
Hence $s'\sigma \in P^{-1}P\overline{K} \cap L(\textbf{G}) =
\overline{K}$ by normality of $\overline{K}$.

For (\ref{eq:c-obs2}), let $s \in K$, $s' \in \overline{C} \cap
L_m(\textbf{G})$; we will prove $s' \in K$. That $s \in K$ implies
$Ps \in PK$; thus $Ps' \in PK$, i.e. $s' \in P^{-1}PK$. Therefore
$s' \in P^{-1}PK \cap L_m(\textbf{G}) = K$ by normality of $K$.
\hfill $\square$

\begin{figure}[!t]
  \centering
  \includegraphics[width=0.48\textwidth]{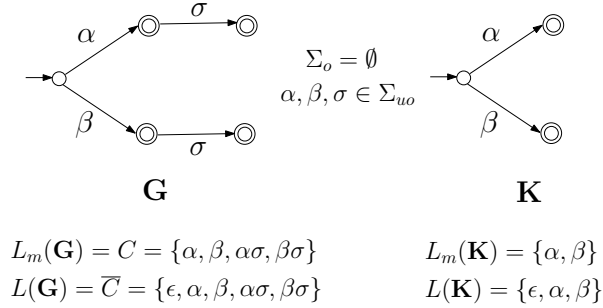}
  \caption{$L_m(\textbf{K})$ is relatively observable but not normal. In $L(\textbf{K})$ all three strings are lookalike;
  it is easily verified that $L_m(\textbf{K})$ is $\overline{C}$-observable. To see that $L_m(\textbf{K})$ is not $(L_m(\textbf{G}),P)$-normal,
  calculate $P^{-1}PL_m(\textbf{K}) = P^{-1}(\epsilon) = \Sigma^*$. Thus $P^{-1}PL_m(\textbf{K}) \cap L_m(\textbf{G}) = L_m(\textbf{G}) \varsupsetneqq L_m(\textbf{K})$.
  A similar calculation yields that $L(\textbf{K})$ is not $(L(\textbf{G}),P)$-normal.}
  \label{fig:c-obs_norm}
\end{figure}

In the proof we note that $\overline{K}$ being
$(L(\textbf{G}),P)$-normal implies condition (i) of relative
observability, and independently $K$ being
$(L_m(\textbf{G}),P)$-normal implies condition (ii). The reverse
statement of Proposition~\ref{prop:c-obs_norm} need not be true; an
example is displayed in Fig.~\ref{fig:c-obs_norm}.

In Section~\ref{sec_examples}, we will see examples where the
supremal relatively observable controlled behavior is strictly
larger than the supremal normal counterpart. This is due exactly to
the distinction as to whether or not one may disable controllable
events that are unobservable.

\begin{figure}[!t]
  \centering
  \includegraphics[width=0.48\textwidth]{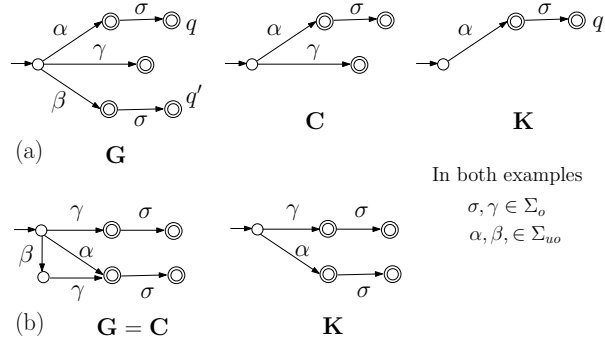}
  \caption{Comparison with \cite{TakaiUshio:03}. In (a), $L_m(\textbf{K})$ is $L({\bf C})$-observable; but ${\bf K}$ is
  not observable in the sense of \cite{TakaiUshio:03}, because state pair $(q,q')$ with $q$ in ${\bf K}$ and $q'$ not in ${\bf K}$
  violates the observability condition in \cite{TakaiUshio:03}. In (b), $\textbf{K}$ is observable in the sense of \cite{TakaiUshio:03};
  but $L_m(\textbf{K})$ is not $L({\bf C})$-observable, because $\gamma \sigma \in L(\textbf{K})$, $\beta \gamma \in L(\textbf{C})$,
  $\beta \gamma \sigma \in L(\textbf{G})$, $P(\gamma)=P(\beta \gamma)$, but $\beta \gamma \sigma \notin L(\textbf{K})$.}
  \label{fig:c-obs_takai}
\end{figure}

We note that \cite{TakaiUshio:03} reported an observability property
which is also weaker than normality. The observability condition in
\cite{TakaiUshio:03} is formulated in a generator form, which is
preserved under a particularly-defined union operation of ``strict
subautomata''. This automata union does not correspond to
language/set union, and hence the reported observability might not
be preserved under set union. In addition, the observability
condition in \cite{TakaiUshio:03} requires checking all state pairs
$(q,q')$ reached by lookalike strings in the whole state set $Q$ of
$\textbf{G}$.  This corresponds to checking all lookalike string
pairs in $L(\textbf{G})$; in this sense, our relative observability
is weaker with the ambient language $\overline{C} \subseteq
L(\textbf{G})$. One such example is provided in
Fig.~\ref{fig:c-obs_takai}(a). This point is also illustrated, when
combined with controllability, in the Guideway example in
Section~\ref{sec_guideway}. However, the reverse case is also
possible, as displayed in Fig.~\ref{fig:c-obs_takai}(b).

\subsection{The supremal relatively observable
sublanguage} \label{subsec:relobs-sup}

First, an arbitrary union of relatively observable languages is
again relatively observable.

\begin{prop} \label{prop:c-obs_union}
Let $K_i \subseteq C$, $i \in I$ (some index set), be
$\overline{C}$-observable. Then $K = \bigcup\{K_i \ |\ i \in I\}$ is
also $\overline{C}$-observable.
\end{prop}

\emph{Proof.} Let $s,s' \in \Sigma^*$ and $Ps = Ps'$.  We must show
that both (\ref{eq:c-obs1}) and (\ref{eq:c-obs2}) hold for $K$.

For (\ref{eq:c-obs1}), let $\sigma \in \Sigma$, $s \sigma \in
\overline{K}$, $s' \in \overline{C}$, and $s' \sigma \in
L(\textbf{G})$; it will be shown that $s' \sigma \in \overline{K}$.
Since $\overline{K} = \overline{\bigcup K_i} = \bigcup
\overline{K_i}$, there exists $j \in I$ such that $s \sigma \in
\overline{K_j}$. But $K_j$ is $\overline{C}$-observable, which
yields $s' \sigma \in \overline{K_j}$. Hence $s' \sigma \in \bigcup
\overline{K_i} = \overline{K}$.

For (\ref{eq:c-obs2}), let $s \in K$, $s' \in \overline{C} \cap
L_m(\textbf{G})$; we will prove $s' \in K$. That $s \in K = \bigcup
K_i$ implies that there exists $j \in I$ such that $s \in K_j$.
Since $K_j$ is $\overline{C}$-observable, we have $s' \in K_j$.
Therefore $s' \in \bigcup K_i = K$.

\hfill $\square$

\begin{figure}[!t]
  \centering
  \includegraphics[width=0.48\textwidth]{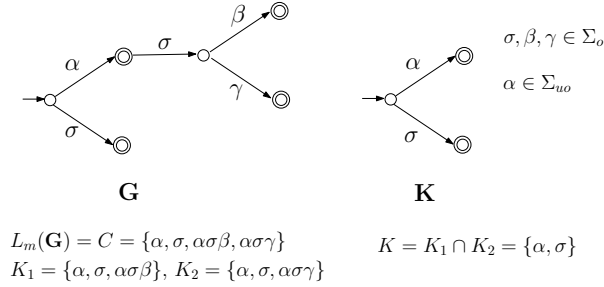}
  \caption{The intersection of two relatively observable languages is not relatively observable. It is easily verified that
  both $K_1$ and $K_2$ are $\overline{C}$-observable. Their intersection $K$, however, is not: let $s=\epsilon$ and $s'=\alpha$; then $Ps=Ps'$,
  $s\sigma \in \overline{K}$, $s' \in \overline{C}$, $s'\sigma \in L({\bf G})$, but $s'\sigma \notin \overline{K}$. Thus condition~(\ref{eq:c-obs1})
  of relative observability is violated.}
  \label{fig:c-obs_intersection}
\end{figure}

While relative observability is closed under arbitrary unions, it is
generally not closed under intersections.
Fig.~\ref{fig:c-obs_intersection} provides an example for which the
intersection of two $\overline{C}$-observable sublanguages is
\emph{not} $\overline{C}$-observable.

Whether or not $K \subseteq C$ is $\overline{C}$-observable, write
\begin{align} \label{eq:c-obs-family}
\mathcal {O}(K, C) := \{ K' \subseteq K \ |\ K' \mbox{ is
$\overline{C}$-observable} \}
\end{align}
for the family of $\overline{C}$-observable sublanguages of $K$. The
discussion above on unions and intersections of relatively
observable languages shows that $\mathcal {O}(K,C)$ is an upper
semilattice of the lattice of sublanguages of $K$, with respect to
the partial order ($\subseteq$).\footnote{For lattice theory refer
to e.g. \cite{DavPri:90},\cite[Chapter~1]{SCDES}.} Note that the
empty language $\emptyset$ is trivially $\overline{C}$-observable,
thus a member of $\mathcal {O}(K,C)$. By
Proposition~\ref{prop:c-obs_union} we derive that $\mathcal
{O}(K,C)$ has a unique supremal element sup$\mathcal {O}(K,C)$ given
by
\begin{align} \label{eq:sup-c-obs}
\mbox{sup}\mathcal {O}(K,C) := \bigcup\{K' \ |\ K' \in \mathcal
{O}(K,C)\}.
\end{align}
This is the supremal $\overline{C}$-observable sublanguage of $K$.
We state these important facts about $\mathcal {O}(K,C)$ in the
following.

\begin{thm} \label{thm:sup-c-obs}
Let $K \subseteq C$. The set $\mathcal {O}(K,C)$ is nonempty, and
contains its supremal element sup$\mathcal {O}(K,C)$ in
(\ref{eq:sup-c-obs}).
\end{thm}


For (\ref{eq:sup-c-obs}), of special interest is when the ambient
language is set to equal $\overline{K}$:
\begin{align} \label{eq:sup-k-obs}
\mbox{sup}\mathcal {O}(K) := \bigcup\{K' \ |\ K' \in \mathcal
{O}(K)\},\ \mbox{ where } \mathcal {O}(K) := \{ K' \subseteq K \ |\
K' \mbox{ is $\overline{K}$-observable} \}
\end{align}

\begin{prop} \label{prop:sup-k-obs}
For $K \subseteq C \subseteq L_m({\bf G})$, it holds that
$\mbox{sup}\mathcal {O}(K,C) \subseteq \mbox{sup}\mathcal {O}(K)$.
\end{prop}
\emph{Proof.} For each $K' \subseteq K$, it follows from
Definition~\ref{defn:c-obs} that if $K'$ is
$\overline{C}$-observable, then $K'$ is also
$\overline{K}$-observable. Hence $\mathcal {O}(K,C) \subseteq
\mathcal {O}(K)$, and sup$\mathcal {O}(K,C) \subseteq$ sup$\mathcal
{O}(K)$. \hfill $\square$

Proposition~\ref{prop:sup-k-obs} shows that sup$\mathcal {O}(K)$ is
the largest relatively observable sublanguage of $K$, given all
choices of the ambient language. It is therefore of particular
interest in characterizing and computing sup$\mathcal {O}(K)$. We do
so in the next section using a generator-based approach.


%% file: sec_relobs-gen.tex
\section{Generator-Based Computation of sup$\mathcal {O}(K)$} \label{sec_relobs-gen}

In this section we design an algorithm that computes the supremal
relatively observable sublanguage sup$\mathcal {O}(K)$ of a given
language $K$. This algorithm has two new mechanisms that distinguish
it from those computing the supremal normal sublanguage (e.g.
\cite{ChoMarcus:89,BrandtLin:90,KomSch:05}): First, compared to
\cite{ChoMarcus:89,BrandtLin:90,KomSch:05}, the algorithm embeds a
more intricate, `fine-grained' procedure (to be stated precisely
below) for processing \emph{transitions} of the generators involved;
this new procedure is needed because relative observability is
weaker than normality, and thus generally requires fewer transitions
to be removed. Second, the algorithm keeps track of strings in the
ambient language $\overline{K}$, as required by the relative
observability conditions; by contrast, this is simply not an issue
in \cite{ChoMarcus:89,BrandtLin:90,KomSch:05} for the normality
computation.


\subsection{Setting}

Consider a nonblocking generator $\textbf{G} = (Q, \Sigma, \delta,
q_0, Q_m)$ as in (\ref{eq:generator}) with regular languages
$L_m(\textbf{G})$ and $L(\textbf{G})$, and a natural projection
$P:\Sigma^* \rightarrow \Sigma^*_o$ with $\Sigma_o \subseteq
\Sigma$.  Let $K$ be an arbitrary regular sublanguage of
$L_m(\textbf{G})$. Then $K$ can be represented by a finite-state
generator ${\bf K} = (Y, \Sigma, \eta, y_0, Y_m)$; that is,
$L_m({\bf K}) = K$ and $L({\bf K}) = \overline{K}$. For simplicity
we assume ${\bf K}$ is nonblocking, i.e.
$\overline{L_m(\textbf{K})}=L(\textbf{K})$. Denote by $n,m$
respectively the number of states and transitions of ${\bf K}$, i.e.
\begin{equation} \label{eq:nm}
\begin{split}
&n := |Y| \\
&m:= |\eta|=|\{(y,\sigma,\eta(y,\sigma)) \in Y \times \Sigma \times
Y \ |\ \eta(y,\sigma)! \}|.
\end{split}
\end{equation}
We introduce

\noindent \emph{Assumption~1.} $(\forall s,t \in L({\bf K}))\
\eta(y_0,s)=\eta(y_0,t) \Rightarrow \delta(q_0,s)=\delta(q_0,t)$.

\noindent If the given ${\bf K}$ does not satisfy Assumption~1, form
the following \emph{synchronous product} (\cite{SCDES,CaLa:07})
\begin{align} \label{eq:syncprod}
{\bf K} || {\bf G} = (Y \times Q, \Sigma, \eta \times \delta,
(y_0,q_0), Y_m \times Q_m)
\end{align}
where $\eta \times \delta : Y \times Q \times \Sigma \rightarrow Y
\times Q$ is given by
\begin{align*}
(\eta \times \delta) \big( (y,q),\sigma \big) = \left\{
                                                \begin{array}{ll}
                                                  \big( \eta(y,\sigma),\delta(q,\sigma) \big), & \hbox{if $\eta(y,\sigma)!$ \& $\delta(q,\sigma)!$;} \\
                                                  \mbox{undefined}, & \hbox{otherwise.}
                                                \end{array}
                                              \right.
\end{align*}
It is easily checked (e.g. \cite[Section~2.3.3]{CaLa:07}) that
$L({\bf K} || {\bf G})=L({\bf K}) \cap L({\bf G})=L({\bf K})$,
$L_m({\bf K} || {\bf G})=L_m({\bf K}) \cap L_m({\bf G})=L_m({\bf
K})$, and for every $s,t \in L({\bf K} || {\bf G})$ if $(\eta \times
\delta) \big( (y_0,q_0),s \big) = (\eta \times \delta) \big(
(y_0,q_0),t \big)$, then $\delta(q_0,s)=\delta(q_0,t)$. Namely ${\bf
K} || {\bf G}$ satisfies Assumption~1. Therefore, replacing ${\bf
K}$ by the synchronous product ${\bf K} || {\bf G}$ always makes
Assumption~1 hold.

Now if for some $s \in L({\bf K})$ a string $Ps \in PL({\bf K})$ is
observed, then the ``uncertainty set'' of states which $s$ may reach
in ${\bf K}$ is
\begin{align} \label{eq:uncer}
U(s) := \{ \eta(y_0,s') \ | \ s' \in L({\bf K}), Ps' = Ps \}
\subseteq Y.
\end{align}
If two strings have the same uncertainty set, then the following is
true.

\begin{lem} \label{lem:uncer}
Let $s,t \in L({\bf K})$ be such that $U(s)=U(t)$.  If $s' \in
L({\bf K})$ looks like $s$, i.e. $Ps'=Ps$, then there exists $t' \in
L({\bf K})$ such that $Pt'=Pt$ and $\eta(y_0,t')=\eta(y_0,s')$.
\end{lem}

\emph{Proof.} Since $s' \in L({\bf K})$ and $Ps'=Ps$, by
(\ref{eq:uncer}) we have $\eta(y_0,s') \in U(s)$.  Then it follows
from $U(s) = U(t)$ that $\eta(y_0,s') \in U(t)$, and hence there
exists $t' \in L({\bf K})$ such that $Pt'=Pt$ and
$\eta(y_0,t')=\eta(y_0,s')$. \hfill $\square$

We further adopt

\noindent \emph{Assumption~2.}
\begin{align} \label{eq:norm-gen}
(\forall s,t \in L({\bf K}))\ \eta(y_0,s)=\eta(y_0,t) \Rightarrow
U(s)=U(t).
\end{align}
Assumption~2 requires that any two strings reaching the same state
of ${\bf K}$ must have the same uncertainty set.  This requirement
is equivalent to the ``normal automaton'' condition in
\cite{ChoMarcus:89,TakaiUshio:03}, which played a key role in their
algorithms.  In case the given ${\bf K}$ does not satisfy
(\ref{eq:norm-gen}), a procedure is presented in
\cite[Appendix~A]{TakaiUshio:03} which makes Assumption~2 hold.
Essentially, the procedure consists of two steps: first, construct a
deterministic generator ${\bf PK}$ with event set $\Sigma_o$
obtained by the \emph{subset construction} such that $L_m({\bf
PK})=PL_m({\bf K})$ and $L({\bf PK})=PL({\bf K})$ (e.g.
\cite[Section~2.5]{SCDES}). The subset construction ensures that if
two strings $Ps,Pt$ reach the same state in ${\bf PK}$, then
$U(s)=U(t)$. The state size of ${\bf PK}$ is at worst exponential in
that of ${\bf K}$. Second, form the synchronous product ${\bf K} ||
{\bf PK}$ as in (\ref{eq:syncprod}), so that $L({\bf K} || {\bf
PK})=L({\bf K}) \cap P^{-1}PL({\bf K})=L({\bf K})$ and $L_m({\bf K}
|| {\bf PK})=L_m({\bf K}) \cap P^{-1}PL_m({\bf K})=L_m({\bf K})$.
Therefore, replacing ${\bf K}$ by ${\bf K} || {\bf PK}$ always makes
Assumption~2 hold. Like Assumption~1, Assumption~2 entails no loss
of generality.

Let Assumptions~1 and 2 hold.  We present an algorithm which
produces a finite sequence of generators
\begin{align} \label{eq:gen-chain}
({\bf K} =) {\bf K}_0,\ \ {\bf K}_1,\ \ \cdots,\ \ {\bf K}_N
\end{align}
with ${\bf K}_i = (Y_i, \Sigma, \eta_i, y_0, Y_{m,i})$, $i \in
[0,N]$, and a corresponding finite descending chain of languages
\begin{align*}
(L_m({\bf K}) =) L_m({\bf K}_0) \supseteq L_m({\bf K}_1) \supseteq
\cdots \supseteq L_m({\bf K}_N)
\end{align*}
such that $L_m({\bf K}_N) =$ sup$\mathcal {O}(K)$ in
(\ref{eq:sup-c-obs}) with the ambient language $\overline{K}$.  If
$K$ is observable (in the standard sense), then $N=0$.


\subsection{Observational consistency}

Given ${\bf K}_i = (Y_i, \Sigma, \eta_i, y_0, Y_{m,i})$, $i \in
[0,N]$, suppose $\overline{L_m({\bf K}_i)}=L({\bf K}_i)$, namely
${\bf K}_i$ is nonblocking. We need to check whether or not
$L_m({\bf K}_i)$ is $\overline{K}$-observable. To this end, we
introduce a generator-based condition, called \emph{observational
consistency}. We proceed in two steps. First, let
\begin{align} \label{eq:full-gen}
{\bf \tilde{K}}_i = (\tilde{Y}_i, \Sigma, \tilde{\eta}_i, y_0,
Y_{m,i})
\end{align}
where $\tilde{Y}_i = Y_i \cup \{y_d\}$, with the \emph{dump state}
$y_d \notin Y_i$, and $\tilde{\eta}_i$ is an extension of $\eta_i$
which is fully defined on $\tilde{Y}_i \times \Sigma$, i.e.
\begin{align} \label{eq:dump-state}
\tilde{\eta}_i(y_0, s)=\left\{
                 \begin{array}{ll}
                   \eta_i(y_0, s), & \hbox{if $s \in L({\bf K}_i)$;} \\
                   y_d, & \hbox{if $s \in \Sigma^* - L({\bf K}_i)$.}
                 \end{array}
               \right.
\end{align}
Clearly, the closed and marked languages of ${\bf \tilde{K}}_i$
satisfy $L({\bf \tilde{K}}_i)=\Sigma^*$ and $L_m({\bf
\tilde{K}}_i)=L_m({\bf K}_i)$.

Second, for each $s \in \Sigma^*$ define a set $T_i(s)$ of state
pairs in $\textbf{G}$ and ${\bf \tilde{K}}_i$ by
\begin{align} \label{eq:T(s)}
T_i(s) := \{ (q,y) \in Q \times \tilde{Y}_i \ |\ (\exists s') Ps'
=Ps, q=\delta(q_0,s'), y=\tilde{\eta}_i(y_0,s'), \eta(y_0,s')! \}.
\end{align}
Thus, a pair $(q,y) \in T_i(s)$ if $q \in Q$ and $y \in \tilde{Y}_i$
are reached by a common string $s'$ that looks like $s$, and this
$s'$ is in $L(\textbf{K})$, namely the ambient $\overline{K}$,
because $\eta(y_0,s')!$.  This $\eta(y_0,s')!$ is the key to
tracking strings in the ambient $\overline{K}$.

\noindent \emph{Remark 1.} If one aims to compute
$\mbox{sup}\mathcal {O}(K,C)$ in (\ref{eq:sup-c-obs}) instead of the
largest $\mbox{sup}\mathcal {O}(K)$ in (\ref{eq:sup-k-obs}) (largest
in the sense of Proposition~\ref{prop:sup-k-obs}), for some ambient
language $C$ satisfying $K \subseteq C \subseteq L_m({\bf G})$, then
replace $T_i(s)$ in (\ref{eq:T(s)}) by
\begin{align} \label{eq:TC(s)}
T^C_i(s) := \{ (q,y) \in Q \times \tilde{Y}_i \ |\ (\exists s') Ps'
=Ps, q=\delta(q_0,s'), y=\tilde{\eta}_i(y_0,s'), \eta^C(y_0,s')! \}.
\end{align}
where $\eta^C$ is the transition function of the generator ${\bf C}$
with $L_m({\bf C})=C$ and $L({\bf C})=\overline{C}$. The rest
follows similarly by using $T^C_i(s)$.

\begin{defn} \label{defn:obs-consis}
We say that $T_i(s)$ is \emph{observationally consistent} (with
respect to ${\bf G}$ and ${\bf \tilde{K}}_i$) if for all
$(q,y),(q',y') \in T_i(s)$ there holds
\begin{align} \label{eq:obs-consis1}
&(\forall \sigma \in \Sigma)\ \tilde{\eta}_i(y, \sigma) \neq y_d,
\delta(q',\sigma)! \Rightarrow \tilde{\eta}_i(y',\sigma) \neq y_d \\
& q' \in Q_m, y \in Y_{m,i} \Rightarrow y' \in Y_{m,i}.
\label{eq:obs-consis2}
\end{align}
\end{defn}
Note that if $T_i(s)$ has only one element, then it is trivially
observationally consistent. Let
\begin{align} \label{eq:T}
\mathcal {T}_i := \{ T_i(s) \ |\ s \in \Sigma^*, |T_i(s)| \geq 2\}.
\end{align}
Then $|\mathcal {T}_i| \leq 2^{|Q| \cdot (|\tilde{Y}_i|)} \leq
2^{|Q| \cdot (n+1)}$, which is finite. The following result states
that checking $\overline{K}$-observability of $L_m({\bf K}_i)$ is
equivalent to checking observational consistency of all state pairs
in each of the $T_i$ occurring in $\mathcal {T}_i$.

\begin{lem}\label{lem:obs-consis}
$L_m({\bf K}_i)$ is $\overline{K}$-observable if and only if for
every $T \in \mathcal {T}_i$, $T$ is observationally consistent with
respect to ${\bf G}$ and ${\bf \tilde{K}}_i$.
\end{lem}

\emph{Proof.} (If) Let $s,s' \in \Sigma^*$ and $Ps = Ps'$.  We must
show that both (\ref{eq:c-obs1}) and (\ref{eq:c-obs2}) hold for
$L_m({\bf K}_i)$.

For (\ref{eq:c-obs1}), let $\sigma \in \Sigma$, $s \sigma \in L({\bf
K}_i)$, $s' \in \overline{K}$, and $s' \sigma \in L(\textbf{G})$; it
will be shown that $s' \sigma \in L({\bf K}_i)$. According to
(\ref{eq:T(s)}) and (\ref{eq:dump-state}), the two state pairs
$(\delta(q_0,s),\tilde{\eta}_i(y_0,s)),
(\delta(q_0,s'),\tilde{\eta}_i(y_0,s'))$ belong to $T(s)$. Now
$s\sigma \in L({\bf K}_i)$ implies
$\tilde{\eta}_i(\tilde{\eta}_i(y_0,s), \sigma) \neq y_d$ (by
(\ref{eq:dump-state})), and $s'\sigma \in L(\textbf{G})$ implies
$\delta(\delta(q_0,s'), \sigma)!$. Since $T(s)$ is observationally
consistent, by (\ref{eq:obs-consis1}) we have
$\tilde{\eta}_i(\tilde{\eta}_i(y_0,s'),\sigma) \neq y_d$. Then it
follows from (\ref{eq:dump-state}) that $s'\sigma \in L({\bf K}_i)$.

For (\ref{eq:c-obs2}), let $s \in L_m({\bf K}_i)$, $s' \in
\overline{K} \cap L_m(\textbf{G})$; we will prove $s' \in L_m({\bf
K}_i)$. Again $(\delta(q_0,s),\tilde{\eta}_i(y_0,s))$,
$(\delta(q_0,s'),\tilde{\eta}_i(y_0,s')) \in T(s)$ according to
(\ref{eq:T(s)}) and (\ref{eq:dump-state}). Now $s \in L_m({\bf
K}_i)=L_m({\bf \tilde{K}}_i)$ implies $\tilde{\eta}_i(y_0,s) \in
Y_{m,i}$, and $s' \in L_m(\textbf{G})$ implies $\delta(q_0,s') \in
Q_m$. Since $T(s)$ is observationally consistent, by
(\ref{eq:obs-consis2}) we have $\tilde{\eta}_i(y_0,s') \in Y_{m,i}$,
i.e. $s' \in L_m({\bf \tilde{K}}_i)=L_m({\bf K}_i)$.

(Only if) Let $T \in \mathcal {T}_i$, and $(q,y),(q',y') \in T$
corresponding respectively to some $s$ and $s'$ with $Ps=Ps'$. We
must show that both (\ref{eq:obs-consis1}) and
(\ref{eq:obs-consis2}) hold.

For (\ref{eq:obs-consis1}), let $\sigma \in \Sigma$,
$\tilde{\eta}_i(y,\sigma) \neq y_d$, and $\delta(q',\sigma)!$. It
will be shown that $\tilde{\eta}_i(y',\sigma) \neq y_d$.  Now $(q,y)
\in T$ and $\tilde{\eta}_i(y,\sigma) \neq y_d$ imply $s\sigma \in
L({\bf K}_i)$ (by (\ref{eq:dump-state})); $(q',y') \in T$ and
$\delta(q',\sigma)!$ imply $s' \in \overline{K}$ and $s'\sigma \in
L(\textbf{G})$.  Since $L_m({\bf K}_i)$ is
$\overline{K}$-observable, by (\ref{eq:c-obs1}) we have $s'\sigma
\in L({\bf K}_i)$, and therefore $\tilde{\eta}_i(y',\sigma) \neq
y_d$.

Finally for (\ref{eq:obs-consis2}), let $y \in Y_{m,i}$, $q' \in
Q_m$. We will show $y' \in Y_{m,i}$. From $(q,y) \in T$ and $y \in
Y_{m,i}$, $s \in L_m({\bf \tilde{K}}_i)=L_m(\textbf{K}_i)$; from
$(q',y') \in T$ and $q' \in Q_m$, $s' \in \overline{K} \cap
L_m(\textbf{G})$. Since $L_m({\bf K}_i)$ is
$\overline{K}$-observable, by (\ref{eq:c-obs2}) we have $s' \in
L_m(\textbf{K}_i)=L_m({\bf \tilde{K}}_i)$, i.e. $y' \in Y_{m,i}$.
\hfill $\square$

If there is $T \in \mathcal {T}_i$ that fails to be observationally
consistent, then there exist state pairs $(q,y),(q',y') \in T$ such
that either (\ref{eq:obs-consis1}) or (\ref{eq:obs-consis2}) or both
are violated. Define two sets $R_T$ and $M_T$ as follows:
\begin{align}
R_T &:=\bigcup_{\sigma \in \Sigma} \{ (y,\sigma,\eta_i(y,\sigma)) \
|\ \eta_i(y,\sigma)! \ \&\ (\exists (q',y') \in T)(
\delta(q',\sigma)! \
\&\ \tilde{\eta}_i(y',\sigma)=y_d ) \} \label{eq:RT}\\
M_T &:=\{y \in Y_{m,i} \ |\ (\exists (q',y') \in T)\ q' \in Q_m \
\&\ y' \notin Y_{m,i}  \}\label{eq:MT}
\end{align}
Thus $R_T$ is a collection of transitions of ${\bf K}_i$, each
having corresponding state pairs $(q,y),(q',y') \in T$ that violate
(\ref{eq:obs-consis1}), while $M_T$ is a collection of marker states
of ${\bf K}_i$, each having corresponding state pairs that violate
(\ref{eq:obs-consis2}). To make $T$ observationally consistent, all
transitions in $R_T$ have to be removed, and all states in $M_T$
unmarked. These constitute the main steps in the algorithm below.

\subsection{Algorithm}

We now present an algorithm which computes sup$\mathcal {O}(K)$ in
(\ref{eq:sup-c-obs}).

\emph{Algorithm 1:} Input $\textbf{G}=(Q,\Sigma,\delta,q_0,Q_m)$,
$\textbf{K}=(Y, \Sigma, \eta, y_0, Y_m)$, and $P:\Sigma^*
\rightarrow \Sigma^*_o$.

\noindent 1. Set $\textbf{K}_0 = (Y_0, \Sigma, \eta_0, y_0,
Y_{m,0})=\textbf{K}$, namely $Y_0=Y$, $Y_{m,0}=Y_m$, and
$\eta_0=\eta$.

\noindent 2. For $i \geq 0$, calculate $\mathcal {T}_i$ as in
(\ref{eq:T}) and (\ref{eq:T(s)}) based on $\textbf{G}$,
$\textbf{K}$, ${\bf \tilde{K}}_i=(\tilde{Y}_i, \Sigma,
\tilde{\eta}_i, y_0, Y_{m,i})$ in (\ref{eq:full-gen}), and $P$.

\noindent 3. For each $T \in \mathcal {T}_i$, check if $T$ is
observationally consistent with respect to ${\bf G}$ and ${\bf
\tilde{K}}_i$ (i.e. check if conditions (\ref{eq:obs-consis1}) and
(\ref{eq:obs-consis2}) are satisfied for all $(q,y),(q',y') \in T$):

If every $T \in \mathcal {T}_i$ is observationally consistent with
respect to ${\bf G}$ and ${\bf \tilde{K}}_i$, then go to Step~4
below.  Otherwise, let
\begin{align}
R_i &:= \bigcup_{T \in \mathcal {T}_i} R_T,\ \mbox{where $R_T$ is defined in (\ref{eq:RT})} \label{eq:Ri}\\
M_i &:= \bigcup_{T \in \mathcal {T}_i} M_T,\ \mbox{where $M_T$ is
defined in (\ref{eq:MT})} \label{eq:Mi}
\end{align}
and set\footnote{Here $\eta_i,\eta'_i$ denote the corresponding sets
of transition triples in $Y_i \times \Sigma \times Y_i$.}
\begin{align} \eta'_i &:= \eta_i - R_i \label{eq:eta'i}\\
Y'_{m,i} &:= Y_{m,i} - M_i.  \label{eq:Ym'i}
\end{align}
Let $\textbf{K}_{i+1} = (Y_{i+1}, \Sigma, \eta_{i+1}, y_0,
Y_{m,i+1}) =$ trim$((Y_{i}, \Sigma, \eta'_{i}, y_0, Y'_{m,i}))$,
where trim$(\cdot)$ removes all non-reachable and non-coreachable
states and corresponding transitions of the argument generator. Now
advance $i$ to $i+1$, and go to Step~2.

\noindent 4. Output $\textbf{K}_{N} := {\bf K}_i$.

Algorithm~1 has two new mechanisms as compared to those computing
the supremal normal sublanguage (e.g.
\cite{ChoMarcus:89,BrandtLin:90,KomSch:05}).  First, the mechanism
of the normality algorithms in
\cite{ChoMarcus:89,BrandtLin:90,KomSch:05} is essentially this: If a
transition $\sigma$ is removed from state $y$ of ${\bf \tilde{K}}_i$
reached by some string $s$, then remove $\sigma$ from all states
$y'$ reached by a lookalike string $s'$, i.e. $Ps=Ps'$. (In fact if
$\sigma$ is unobservable, then all the states $y$ and $y'$ as above
are removed.) This (all or nothing) mechanism generally causes,
however, `overkill' of transitions (i.e. removing more transitions
than necessary) in our case of relative observability, because the
latter is weaker than normality and allows more permissive behavior.
Indeed, some $\sigma$ transitions at states $y'$ as above may be
preserved without violating relative observability. Corresponding to
this feature, Algorithm~1 employs a new, fine-grained mechanism: in
Step~3, remove as in (\ref{eq:eta'i}) only those transitions of
${\bf \tilde{K}}_i$ that violate the relative observability
conditions. Moreover, the second new mechanism of Algorithm~1 is
that it keeps track of strings in the ambient language $L({\bf K})$
at each iteration by computing $\mathcal {T}_i$ in (\ref{eq:T}) with
$T_i$ in (\ref{eq:T(s)}) in Step~2 above.  It is these two new
mechanisms that enable Algorithm~1 to compute the supremal
relatively observable sublanguage sup$\mathcal {O}(K)$ in
(\ref{eq:sup-c-obs}).

The two new mechanisms of Algorithm~1 come with an extra
computational cost as compared to the normality algorithms in
\cite{ChoMarcus:89,BrandtLin:90,KomSch:05}. The extra cost is
precisely the computation of $\mathcal {T}_i$ in (\ref{eq:T}), which
is in the worst case exponential in $n$ because $|\mathcal {T}_i|
\leq 2^{(n+1)|Q|}$.  While complexity is an important issue for
practical computation, we shall leave for future research the
problem of finding more efficient alternatives to Algorithm~1. In
our empirical study in Section~\ref{sec_examples}, the supremal
relatively observable sublanguages corresponding to generators with
state size of the order $10^3$ are computed reasonably fast by
Algorithm~1 (see the AGV example).

Algorithm~1 terminates in finite steps: in (\ref{eq:eta'i}), the set
$R_T$ of transitions for every (observationally inconsistent) $T \in
\mathcal {T}_i$ is removed; in (\ref{eq:Ym'i}), the set $M_T$ of
marker states for every (observationally inconsistent) $T \in
\mathcal {T}_i$ is unmarked. At each iteration of Algorithm~1, if at
Step~3 there is an observationally inconsistent $T$, then at least
one of the two sets $R_i$ in (\ref{eq:Ri}) and $M_i$ in
(\ref{eq:Mi}) is nonempty. Therefore at least one transition is
removed and/or one marker state is unmarked. As initially in
$\textbf{K}_0=\textbf{K}$ there are $m$ transitions and $|Y_m|$
$(<n)$ marker states, Algorithm~1 terminates in at most $n+m$
iterations. The complexity of Algorithm~1 is $O((n+m)2^{(n+1)|Q|})$,
because the search ranges $\mathcal {T}_i$ are such that $|\mathcal
{T}_i| \leq 2^{(n+1)|Q|}$. Note that if ${\bf K}$ does \emph{not}
satisfy Assumption~2, we have to replace ${\bf K}$ by ${\bf K}||{\bf
PK}$ and then the complexity of Algorithm~1 is
$O((2^n+m)2^{(2^n+1)|Q|})$.


Note that from $\textbf{K}_{i}$ to $\textbf{K}_{i+1}$ in Step~3
above, for all $s,t \in \Sigma^*$ if $\eta_{i+1}(y_0,s)!$,
$\eta_{i+1}(y_0,t)!$, then $\eta_{i}(y_0,s)!$, $\eta_{i}(y_0,t)!$,
and
\begin{align} \label{eq:eta-relation}
\eta_{i+1}(y_0,s)=\eta_{i+1}(y_0,t) \Rightarrow
\eta_{i}(y_0,s)=\eta_{i}(y_0,t)
\end{align}


Now we state our main result.

\begin{thm} \label{thm:sup-c-obs}
Let Assumptions~1 and 2 hold. Then the output $\textbf{K}_{N}$ of
Algorithm~1 satisfies $L_m(\textbf{K}_{N}) = $ sup$\mathcal {O}(K)$,
the supremal $\overline{K}$-observable sublanguage of $K$.
\end{thm}

\begin{figure}[!t]
  \centering
  \includegraphics[width=0.48\textwidth]{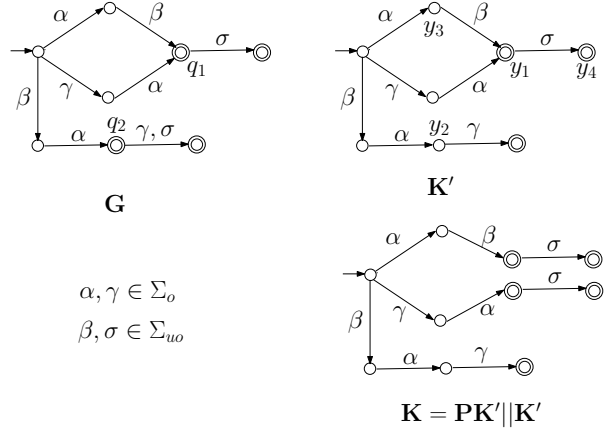}
  \caption{Generator ${\bf K'}$ does not satisfy (\ref{eq:norm-gen}):
  strings $\alpha \beta$ and $\gamma \alpha$ both reach state $y_1$, but $U(\alpha \beta) = \{y_1,y_2,y_3,y_4\} \supsetneqq U(\gamma \alpha) = \{y_1,y_4\}$.
  Now $T(\alpha)$ is not observationally consistent; indeed, two state pairs $(q_1,y_1),(q_2,y_2) \in T(\alpha)$ violate both (\ref{eq:obs-consis1})
  (for transition $\sigma$) and (\ref{eq:obs-consis2}). Applying Algorithm~1 will remove $\sigma$ at $y_1$ and unmark $y_1$, which unintentionally
  removes string $\gamma \alpha \sigma$ and unmarks $\gamma \alpha$. These latter two strings, however, belong to the supremal $\overline{K'}$-observable
  sublanguage. This undesirable situation is avoided in ${\bf K}$ where the strings $\alpha \beta$ and $\gamma \alpha$ are arranged to reach
  different states, and it is easily checked that ${\bf K}$ satisfies (\ref{eq:norm-gen}).}
  \label{fig:norm-gen}
\end{figure}

The condition (\ref{eq:norm-gen}) of Assumption~2 on ${\bf K}$ is
important for Algorithm~1 to generate the supremal relatively
observable sublanguage, because it avoids removing and/or unmarking
a string which is not intended. An illustration is displayed in
Fig.~\ref{fig:norm-gen}.

\begin{figure}[!t]
  \centering
  \includegraphics[width=0.48\textwidth]{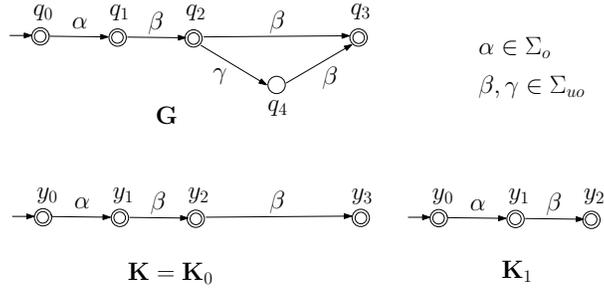}
  \caption{In ${\bf K}_0$, state pairs $(q_1,y_1),(q_2,y_2) \in T(\alpha)$ are observationally consistent, while $(q_2,y_2),(q_4,y_d) \in T(\alpha)$ are not ($y_d$ is the dump state):
  (\ref{eq:obs-consis1}) is violated for transition $\beta$.  Applying Algorithm~1 will remove $\beta$ at $y_2$, and the result is ${\bf K}_1$.
  In ${\bf K}_1$, state pairs $(q_1,y_1),(q_2,y_2) \in T(\alpha)$ become observationally inconsistent: again (\ref{eq:obs-consis1})
  is violated for transition $\beta$. Algorithm~1 needs to be applied again to remove $\beta$ at $y_1$.}
  \label{fig:f-obs-consis}
\end{figure}

Note that removing a transition and/or unmarking a state may destroy
observational consistency of other state pairs.
Fig.~\ref{fig:f-obs-consis} displays such an example. This implies
that all state pairs need to be checked for observational
consistency at each iteration of Algorithm~1.

In addition, just for checking $\overline{C}$-observability of a
given language $K$, with $K \subseteq C$, a polynomial algorithm
(see \cite[Section~3.7]{CaLa:07}, \cite{Tsi:89}) for checking
(standard) observability may be adapted. Indeed, let ${\bf C}$ be a
generator representing $C$, and instead of forming the synchronous
product ${\bf G} || {\bf K} || {\bf K}$ as in
\cite[Section~3.7]{CaLa:07}, we form ${\bf G} || {\bf C} || {\bf
K}$; the rest is similar as may be easily confirmed.


Finally, we provide an example to illustrate the operations involved
in Algorithm 1.

\begin{exmp} \label{ex:supobs-alg}
\begin{figure}[!t]
  \centering
  \includegraphics[width=0.6\textwidth]{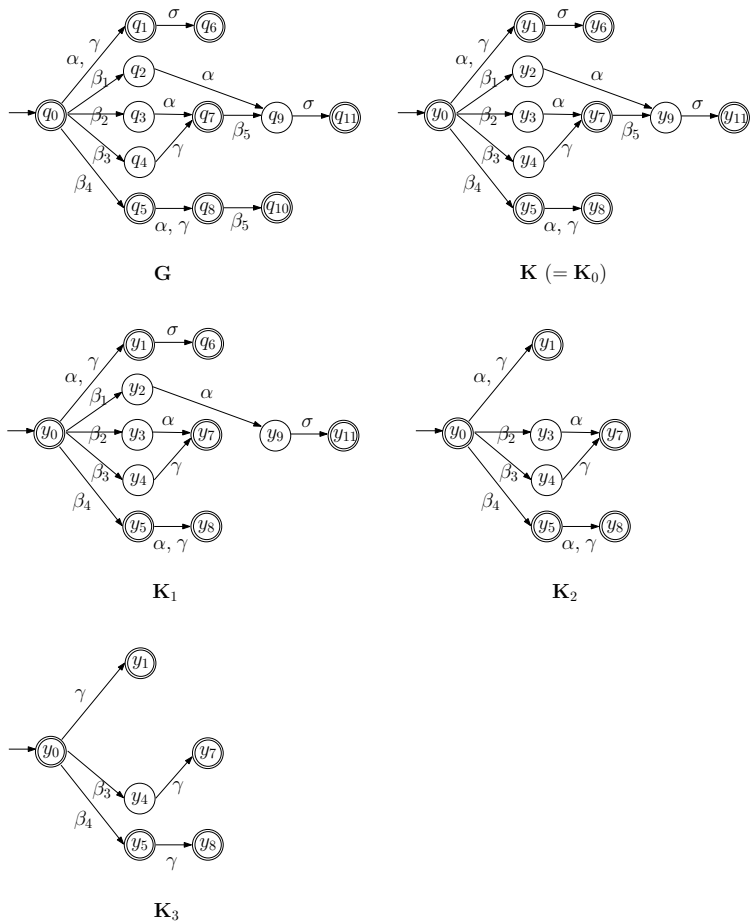}
  \caption{Example illustration of Algorithm 1. Events $\beta_1,...,\beta_5$ are
  unobservable and $\alpha, \gamma, \sigma$ observable.}
  \label{fig:ex_supobs}
\end{figure}

Consider generators ${\bf G}$ and ${\bf K}$ displayed in
Fig.~\ref{fig:ex_supobs}, where events $\beta_1,...,\beta_5$ are
unobservable and $\alpha, \gamma, \sigma$ observable. These events
define a natural projection $P$.
It is easily checked that Assumption~1 holds. Also, in ${\bf K}$, we
have $U(\alpha)=U(\gamma)=\{y_1,y_7,y_8,y_9\}$ and $U(\alpha
\sigma)=U(\gamma \sigma)=\{y_6,y_{11}\}$; thus ${\bf K}$ satisfies
(\ref{eq:norm-gen}) and Assumption~2 holds.

Apply Algorithm~1 with inputs ${\bf G}$, ${\bf K}$, and the natural
projection $P$. Set ${\bf K}_0 = {\bf K}$, and compute $\mathcal
{T}_0 =\{T_1, T_2, T_3\}$ with
\begin{align*}
T_1 &= \{(q_0,y_0),(q_2,y_2),(q_3,y_3),(q_4,y_4),(q_5,y_5)\} \ (=T(\epsilon))\\
T_2 &= \{(q_1,y_1),(q_7,y_7),(q_8,y_8),(q_9,y_9)\} \ (=T(\alpha)=T(\gamma)) \\
T_3 &= \{(q_6,y_6),(q_{11},y_{11})\} \ (=T(\alpha \sigma)=T(\gamma
\sigma)).
\end{align*}
While $T_1,T_3$ are observationally consistent with respect to ${\bf
K}_0$, $T_2$ is not; indeed, $(q_7,y_7),(q_8,y_8)$ violate
(\ref{eq:obs-consis1}) with event $\beta_5$. Thus
$R_0=\{(y_7,\beta_5,y_9)\}$ and $M_0=\emptyset$; the unobservable
transition $(y_7,\beta_5,y_9)$ is removed, which yields a trim
generator ${\bf K}_1$ in Fig.~\ref{fig:ex_supobs}.

The above is the first iteration of Algorithm~1. Next, compute
$\mathcal {T}_1 =\{T_1, T_2, T_3, T_4, T_5\}$ with
\begin{align*}
T_1 &= \{(q_0,y_0),(q_2,y_2),(q_3,y_3),(q_4,y_4),(q_5,y_5)\} \ (=T(\epsilon))\\
T_2 &= \{(q_1,y_1),(q_7,y_7),(q_8,y_8),(q_9,y_d)\} \ (=T(\gamma)) \\
T_3 &= \{(q_1,y_1),(q_7,y_7),(q_8,y_8),(q_9,y_9),(q_9,y_d)\} \ (=T(\alpha)) \\
T_4 &= \{(q_6,y_d),(q_{11},y_{d})\} \ (T(\gamma \sigma))\\
T_5 &= \{(q_6,y_d),(q_{11},y_{11}), (q_{11},y_{d})\} \ (=T(\alpha
\sigma)).
\end{align*}
Note that $T(\alpha) \neq T(\gamma)$ and $T(\alpha \sigma) \neq
T(\gamma \sigma)$ in ${\bf K}_1$, although $T(\alpha) = T(\gamma)$
and $T(\alpha \sigma) = T(\gamma \sigma)$ in ${\bf K}_0$. Now
$T_2$,...,$T_5$ are all observationally \emph{inconsistent} with
respect to ${\bf K}_1$, and $R_1=\{(y_1,\sigma,y_6)\}$,
$M_1=\{y_{11}\}$. Thus removing transition $(y_1,\sigma,y_6)$,
unmarking $y_{11}$, and trimming the result yield ${\bf K}_2$ in
Fig.~\ref{fig:ex_supobs}. This finishes the second iteration of
Algorithm~1.

Compute $\mathcal {T}_3 =\{T_1, T_2\}$ with
\begin{align*}
T_1 &= \{(q_0,y_0),(q_2,y_d),(q_3,y_3),(q_4,y_4),(q_5,y_5)\} \ (=T(\epsilon))\\
T_2 &= \{(q_1,y_1),(q_7,y_7),(q_8,y_8),(q_9,y_d)\} \
(=T(\gamma)=T(\alpha)).
\end{align*}
Here $T_1$ is not observationally consistent, and
$R_3=\{(y_0,\alpha,y_1),(y_3,\alpha,y_7),(y_5,\alpha,y_8)\}$,
$M_3=\emptyset$. Thus removing these three transitions and trimming
the result yield ${\bf K}_3$ in Fig.~\ref{fig:ex_supobs}. This is
the third iteration of Algorithm~1. Now compute $\mathcal {T}_4
=\{T_1, T_2\}$ with
\begin{align*}
T_1 &= \{(q_0,y_0),(q_2,y_d),(q_3,y_d),(q_4,y_4),(q_5,y_5)\} \ (=T(\epsilon))\\
T_2 &= \{(q_1,y_1),(q_7,y_7),(q_8,y_8),(q_9,y_d)\} \ (=T(\gamma)).
\end{align*}
It is easily checked that both $T_1$ and $T_2$ are observationally
consistent with respect to ${\bf K}_3$; by
Lemma~\ref{lem:obs-consis}, $L_m({\bf K}_3)$ is $L({\bf
K})$-observable.  Hence Algorithm~1 terminates after four
iterations, and outputs ${\bf K}_3$. By Theorem~\ref{thm:sup-c-obs},
$L_m({\bf K}_3)$ is in fact the supremal $L({\bf K})$-observable
sublanguage of $L({\bf K})$. By contrast, the supremal normal
sublanguage of $L({\bf K})$ is empty.
\end{exmp}


We now prove Theorem~\ref{thm:sup-c-obs}.

\emph{Proof of Theorem~\ref{thm:sup-c-obs}.} We show
$L_m(\textbf{K}_{N}) = $ sup$\mathcal {O}(K)$. First, it is
guaranteed by Algorithm~1 that for the output $\textbf{K}_{N}$, all
the corresponding $T \in \mathcal {T}_N$ are observationally
consistent; hence Lemma~\ref{lem:obs-consis} implies that
$L_m(\textbf{K}_{N})$ is $\overline{K}$-observable.

It remains to prove that if $K' \in \mathcal {O}(K)$, then $K'
\subseteq L_m(\textbf{K}_{N})$. We proceed by induction on the
iterations $i=0,1,2,...$ of Algorithm~1.  Since $K' \subseteq K =
L_m(\textbf{K})$, we have $K' \subseteq L_m(\textbf{K}_0)$. Suppose
now $K' \subseteq L_m(\textbf{K}_i)$; we show that $K' \subseteq
L_m(\textbf{K}_{i+1})$. Let $w \in K'$; by hypothesis $w \in
L_m(\textbf{K}_i)$.  It will be shown that $w \in
L_m(\textbf{K}_{i+1})$ as well.

First, suppose on the contrary that $w \notin L({\bf K}_{i+1})$.
Since $w \in L({\bf K}_i)$, there exist $t \in \Sigma^*$ and $\sigma
\in \Sigma$ such that $t\sigma \leq w$, $\eta_i(y_0,t)=:y \in Y_i$,
and $(y,\sigma,\eta_i(y,\sigma)) \in R_i$ in (\ref{eq:Ri}). Then
there is $T \in \mathcal {T}_i$ such that
$(y,\sigma,\eta_i(y,\sigma)) \in R_T$ in (\ref{eq:RT}), and $T$ is
\emph{not} observationally consistent ((\ref{eq:obs-consis1}) is
violated). Since $K'$ is $\overline{K}$-observable and $t \in
\overline{K'}$, Lemma~\ref{lem:obs-consis} implies that $T(t)$ is
observationally consistent, and thus $T(t) \neq T$.

Now let $s \in \Sigma^*$ be such that $s \neq t$,
$\eta_i(y_0,s)=\eta_i(y_0,t)=y$, and $T(s)=T$. Then by (\ref{eq:RT})
there exists $(q',y') \in T(s)$ such that $\delta(q',\sigma)!$ and
$\tilde{\eta}_i(y',\sigma)=y_d$. Let $s' \in L({\bf K}_0)=L({\bf
K})$ be such that $Ps=Ps'$, $\delta(q_0,s')=q'$, and
$\tilde{\eta}_i(y_0,s')=y'$.  Whether or not $y'=y_d$, there must
exist $s'_1, u \in \Sigma^*$ such that $s'_1 u = s'$,
$\tilde{\eta}_i(y_0,s'_1) \neq y_d$ (i.e. $\eta_i(y_0,s'_1)!$), and
the following is true: if $u=\epsilon$ then
$\tilde{\eta}_i(y_0,s'_1\sigma) = y_d$; otherwise, for each $u_1 \in
\overline{\{u\}}-\{\epsilon\}$, $\tilde{\eta}_i(y_0,s'_1u_1) = y_d$.
We claim that $u \in \Sigma^*_{uo}$, i.e. an unobservable string.
Otherwise, if there exist $u_1 \leq u$ and $\alpha \in \Sigma_o$
such that $u_1 \alpha \leq u$, then by $Ps=Ps'$ there is $s_1 \leq
s$ such that $s_1\alpha \leq s$ and $Ps_1=P(s'_1u_1)$. Since
$\tilde{\eta}_i(y_0,s'_1u_1\alpha)=y_d$, we have
$(\eta_j(y_0,s_1),\alpha,\eta_j(y_0,s_1 \alpha)) \in R_j$ for some
$j < i$. Hence $s \notin L({\bf K}_i)$, which is contradicting our
choice of $s$ that $\eta_i(y_0,s)=\eta_i(y_0,t)=y$.

Now $u \in \Sigma^*_{uo}$ and $s'_1 u = s'$ imply $Ps'_1=Ps'=Ps$.
Since $\eta_i(y_0,s)=\eta_i(y_0,t)=y$, by repeatedly using
(\ref{eq:eta-relation}) we derive $\eta_0(y_0,s)=\eta_0(y_0,t)=y$.
Then by Assumption~2 and Lemma~\ref{lem:uncer}, there exists $t' \in
L({\bf K}_0)$ such that $Pt=Pt'$ and
$\eta_0(y_0,t')=\eta_0(y_0,s'_1)$. Thus
$\eta_0(y_0,t'u)=\eta_0(y_0,s'_1u)$. It then follows from
Assumption~1 and $\eta_0=\eta$ that
$\delta(q_0,t'u)=\delta(q_0,s'_1u)=q'$ and
$\delta(\delta(q_0,t'u),\sigma)!$. On the other hand,
$\tilde{\eta}_i(\tilde{\eta}_i(y_0,t'u),\sigma)=\tilde{\eta}_i(\tilde{\eta}_i(y_0,s'_1u),\sigma)=y_d$.
Since $P(t'u)=Pt'=Pt$, we have
$(\delta(q_0,t'u),\tilde{\eta}_i(y_0,t'u)) \in T(t)$. This implies
that $T(t)$ is \emph{not} observationally consistent, which
contradicts that $K'$ is $\overline{K}$-observable. Therefore $w \in
L({\bf K}_{i+1})$.

Next, suppose $w \in L({\bf K}_{i+1}) - L_m({\bf K}_{i+1})$. Since
$w \in L_m({\bf K}_i)$, we have $\eta_i(y_0,w) =: y_m$ and $y_m \in
M_i$ in (\ref{eq:Mi}). Then there is $T \in \mathcal {T}_i$ such
that $y_m \in M_T$ in (\ref{eq:MT}), and $T$ is \emph{not}
observationally consistent ((\ref{eq:obs-consis2}) is violated).
Since $K'$ is $\overline{K}$-observable and $w \in K'$,
Lemma~\ref{lem:obs-consis} implies that $T(w)$ is observationally
consistent, and thus $T(w) \neq T$.

Now let $v \in \Sigma^*$ be such that $v \neq w$,
$\eta_i(y_0,v)=\eta_i(y_0,w)=y_m$, and $T(v)=T$. Then by
(\ref{eq:MT}) there exists $(q'_m,y'_m) \in T(v)$ such that $q'_m
\in Q_m$ and $y'_m \notin Y_{m,i}$. Let $v' \in L({\bf K}_0)=L({\bf
K})$ be such that $Pv=Pv'$, $\delta(q_0,v')=q'_m$, and
$\tilde{\eta}_i(y_0,v')=y'_m$.  Whether or not $y'_m=y_d$, by a
similar argument to the one above we derive that there exists $w'$,
with $Pw=Pw'$, such that $\delta(y_0,w')=\delta(y_0,v')=q'_m$,
$\eta_0(y_0,w')=\eta_0(y_0,v')$, and
$\tilde{\eta}_i(y_0,w')=\tilde{\eta}_i(y_0,v')=y'_m \notin Y_{m,i}$.
It follows that $(\delta(q_0,w'),\tilde{\eta}_i(y_0,w')) \in T(w)$.
This implies that $T(w)$ is \emph{not} observationally consistent,
which contradicts that $K'$ is $\overline{K}$-observable. Therefore
$w \in L_m({\bf K}_{i+1})$, and the proof is complete. \hfill
$\square$

\subsection{Polynomial Complexity under $L_m({\bf K})$-Observer}

Given a general natural projection $P : \Sigma^* \rightarrow
\Sigma^*_o$, $\Sigma_o \subseteq \Sigma$, we have seen that the
(worst-case) complexity of Algorithm~1 is exponential in $n$
(defined in (\ref{eq:nm}) as the state size of generator ${\bf K}$).
Does there exist a special class of natural projections $P$ for
which the complexity of Algorithm~1 is polynomial in $n$? In this
section we provide an answer to this question: we identify a
condition on $P$ that suffices to guarantee polynomial complexity in
$n$ of Algorithm~1. Moreover, the condition itself is verified with
polynomial complexity in $n$.

The condition is \emph{$L_m({\bf K})$-observer} \cite{FenWon:08}:
Let ${\bf K} = (Y,\Sigma,\eta,y_0,Y_m)$ ($|Y|=n$) be a finite-state
generator and $P : \Sigma^* \rightarrow \Sigma^*_o$ a natural
projection with $\Sigma_o \subseteq \Sigma$. We say that $P$ is an
$L_m({\bf K})$-observer if
\begin{align} \label{eq:LmObs}
(\forall s \in L({\bf K}), \forall t_o \in \Sigma^*_o)\ (Ps)t_o \in
PL_m({\bf K}) \Rightarrow (\exists t \in \Sigma^*)\ Pt=t_o \ \&\ st
\in L_m({\bf K}).
\end{align}
Thus whenever $Ps$ can be extended to $PL_m({\bf K})$ by an
observable string $t_o$, the underlying string $s$ can be extended
to $L_m({\bf K})$ by a string $t$ with $Pt = t_o$. This condition
plays a key role in nonblocking supervisory control for large-scale
DES [25], and is checkable with polynomial complexity
$|\Sigma|\cdot|Y|^4=|\Sigma| \cdot n^4$ \cite{FenWon:10}. The key
property of $L_m({\bf K})$-observer we use here is the following
fact \cite[Section 6.7]{SCDES}.

\begin{lem} \label{lem:LmObs}
Let ${\bf PK}$ over $\Sigma_o$ be the deterministic generator
obtained by subset construction, with marked language $L_m({\bf
PK})=PL_m({\bf K})$ and closed language $L({\bf PK})=PL({\bf K})$.
If $P : \Sigma^* \rightarrow \Sigma^*_o$ is an $L_m({\bf
K})$-observer, then $|{\bf PK}| \leq n$, where $|{\bf PK}|$ is the
state size of ${\bf PK}$.
\end{lem}

Namely, $P$'s $L_m({\bf K})$-observer property renders the
corresponding subset construction \emph{linear}, which would
generally be exponential.  This is because, when $P$ is an $L_m({\bf
K})$-observer, the corresponding subset construction is equivalent
to a (canonical) reduction of ${\bf K}$ by partitioning its state
set $Y$; the latter results in ${\bf PK}$ with state size no more
than $|Y|=n$ \cite{FenWon:08}.

Now recall ${\bf G}=(Q,\Sigma,\delta,q_0,Q_m)$ with marked language
$L_m({\bf G})$ and closed language $L({\bf G})$, and ${\bf K} =
(Y,\Sigma,\eta,y_0,Y_m)$ ($|Y|=n$, $|\eta|=m$) representing the
regular language $K \subseteq L_m({\bf G})$. We state the main
result of this subsection.

\begin{thm} \label{thm:LmObs}
If $P : \Sigma^* \rightarrow \Sigma^*_o$ is an $L_m({\bf
K})$-observer, then Algorithm~1 has polynomial complexity $|Q|^2
\cdot (n+1)^2 \cdot (n+m) = O(n^3)$.
\end{thm}

\emph{Proof.} For a general natural projection, the exponential
complexity of Algorithm~1 is due to the fact that the sets $\mathcal
{T}_i$ in (\ref{eq:T}), $i \geq 0$, have sizes $|\mathcal {T}_i|
\leq 2^{|Q| \cdot (n+1)}$. We show that $|\mathcal {T}_i| \leq |Q|
\cdot (n+1)$ when $P : \Sigma^* \rightarrow \Sigma^*_o$ is an
$L_m({\bf K})$-observer.

Suppose that $P : \Sigma^* \rightarrow \Sigma^*_o$ is an $L_m({\bf
K})$-observer. First let ${\bf \tilde{K}}$ as in (\ref{eq:full-gen})
be the extension of ${\bf K}$ with a dump state and corresponding
transitions. Thus $|{\bf \tilde{K}}|=n+1$. Consider the synchronous
product ${\bf G} || {\bf \tilde{K}}$ as in (\ref{eq:syncprod}).
Since $L_m({\bf \tilde{K}} || {\bf G})=L_m({\bf K}) \cap L_m({\bf
G})=L_m({\bf K})$, it is easily verified according to
(\ref{eq:LmObs}) that $P$ is also an $L_m({\bf G}||{\bf
\tilde{K}})$-observer. Write ${\bf F}$ for ${\bf G} || {\bf
\tilde{K}}$; then by Lemma~\ref{lem:LmObs}, the generator ${\bf PF}$
over $\Sigma_o$ by applying subset construction to ${\bf G} || {\bf
\tilde{K}}$ is such that $|{\bf PF}| \leq |Q| \cdot |{\bf
\tilde{K}}| = |Q| \cdot (n+1)$.

Now by the definition of $T_i(s)$ in (\ref{eq:T(s)}), $i \geq 0$ and
$s \in \Sigma^*$, we have $T_i(s)=T_i(s')$ whenever $Ps=Ps'$. Hence,
the number of distinct $T_i(s)$ is no more than the state size of
${\bf PF}_i$ obtained by applying subset construction to ${\bf G} ||
{\bf \tilde{K}}_i$. Since $|{\bf \tilde{K}}_i| \leq |{\bf
\tilde{K}}|$, we derive $|{\bf PF}_i| \leq |{\bf PF}| \leq |Q| \cdot
(n+1)$, and therefore
\begin{align*}
|\mathcal {T}_i| \leq |{\bf PF}_i| \leq |Q| \cdot (n+1).
\end{align*}

Finally, since $|T_i(s)| \leq |Q|\cdot|{\bf \tilde{K}}_i|$ for all
$i \geq 0$ and $s \in \Sigma^*$, and Algorithm~1 terminates in at
most $(n+m)$ iterations, we conclude that the complexity of
Algorithm~1 is
\begin{align*}
(n+m) \cdot |\mathcal {T}_i| \cdot |T_i(s)| &\leq (n+m) \cdot (|Q|
\cdot (n+1)) \cdot (|Q| \cdot (n+1))\\
&= |Q|^2 \cdot (n+1)^2 \cdot (n+m) = O(n^3).
\end{align*}
\hfill $\square$

Using Algorithm~1 to compute the supremal relatively observable
sublanguage of a given language $K$, by Theorem~\ref{thm:sup-c-obs}
${\bf G}$ and ${\bf K}$ must satisfy Assumptions 1 and 2. As we have
discussed in Section III.A, Assumption~1 is always satisfied if we
replace ${\bf K}$ by the synchronous product ${\bf G} || {\bf K}$,
which is at most of state size $|Q|\cdot n$.  Assumption~2 is always
satisfied if we replace ${\bf K}$ by ${\bf K} || {\bf PK}$. The
latter has state size at most $n^2$, when the corresponding natural
projection $P : \Sigma^* \rightarrow \Sigma^*_o$ is an $L_m({\bf
K})$-observer.  Therefore, the computation of the supremal
relatively observable sublanguage of a given language $K$ by
Algorithm~1 is of polynomial complexity if $P$ is an $L_m({\bf
K})$-observer.

A procedure of polynomial complexity $O(n^4)$ is available to check
if a given $P$ is an $L_m({\bf K})$-observer \cite{FenWon:10}. If
the check is positive, then by Theorem~\ref{thm:LmObs} we are
assured that the computation of Algorithm~1 is of polynomial
complexity.  In the case that $P$ fails to be an $L_m({\bf
K})$-observer, one may still use Algorithm~1 with the worst-case
exponential complexity.  An alternative in this case is to employ a
polynomial algorithm in \cite{FenWon:10} that extends $P$ to be an
$L_m({\bf K})$-observer by adding more events to $\Sigma_o$.
Thereby polynomial complexity of Algorithm~1 is guaranteed at the
cost of observing more events; this may be helpful in situations
where one has some design freedom in the observable event subset
$\Sigma_o$.


%% file: sec_relobs-con.tex
\section{Supremal Relatively Observable and Controllable Sublanguage} \label{sec_relobs-con}

Consider a plant ${\bf G}$ as in (\ref{eq:generator}) with $\Sigma =
\Sigma_c \dot{\cup} \Sigma_{u}$, where $\Sigma_c$ is the
controllable event subset and $\Sigma_{u}$ the uncontrollable
subset.  A language $K \subseteq L_m({\bf G})$ is
\emph{controllable} (with respect to ${\bf G}$ and $\Sigma_{u}$) if
$\overline{K} \Sigma_u \cap L({\bf G}) \subseteq \overline{K}$. A
\emph{supervisory control} for ${\bf G}$ is any map $V:L({\bf G})
\rightarrow \Gamma$, where $\Gamma:=\{ \gamma \subseteq \Sigma |
\gamma \supseteq \Sigma_u \}$. Then the closed-loop system is
$V/{\bf G}$, with closed behavior $L(V/{\bf G})$ and marked behavior
$L_m(V/{\bf G})$.  Let $\Sigma_o \subseteq \Sigma$ and $P:\Sigma^*
\rightarrow \Sigma^*_o$ be a natural projection. We say $V$ is
\emph{feasible} if $(\forall s, s' \in L({\bf G}))\ P(s)=P(s')
\Rightarrow V(s)=V(s')$, and $V$ is \emph{nonblocking} if
$\overline{L_m(V/{\bf G})} = L(V/{\bf G})$.

It is well known \cite{LinWon:88a} that a feasible nonblocking
supervisory control $V$ exists which synthesizes a nonempty
sublanguage $K \subseteq L_m({\bf G})$ if and only if $K$ is both
controllable and observable.\footnote{Here we let $L_m(V/{\bf
G})=L(V/{\bf G}) \cap K$, namely marking is part of supervisory
control $V$'s action.  In this way we do not need to assume that $K$
is $L_m({\bf G})$-closed, i.e. $K = \overline{K} \cap L_m({\bf G})$
\cite[Section~6.3]{SCDES}.} When $K$ is not observable, however,
there generally does not exist the supremal controllable and
observable sublanguage of $K$.  In this case, the stronger normality
condition is often used instead of observability, so that one may
compute the supremal controllable and normal sublanguage of $K$
\cite{LinWon:88a,Cieslak:88}. With normality ($K$ is
$(L_m(\textbf{G}),P)$-normal and $\overline{K}$ is
$(L(\textbf{G}),P)$-normal), however, no unobservable controllable
event may be disabled; for some applications the resulting
controlled behavior might thus be overly conservative.

This section will present an algorithm which computes, for a given
language $K \subseteq L_m({\bf G})$, a controllable and relatively
observable sublanguage $K_\infty$ that is generally larger than the
supremal controllable and normal sublanguage of $K$.  In particular,
it allows disabling unobservable controllable events.  Being
relatively observable, $K_\infty$ is also observable and
controllable, and thus may be synthesized by a feasible nonblocking
supervisory control.

First, the algorithm which computes the supremal controllable
sublanguage of a given language is reviewed \cite{WonRam:87}. Given
a language $K \subseteq L_m({\bf G})$, whether controllable or not,
write $\mathcal {C}(K) := \{ K' \subseteq K \ |\ K' \mbox{ is
controllable} \}$ for the family of controllable sublanguages of
$K$. Then $\mathcal {C}(K)$ is nonempty ($\emptyset$ belongs) and
has a unique supremal element sup$\mathcal {C}(K):= \bigcup\{K' \ |\
K' \in \mathcal {C}(K)\}$ \cite{SCDES}.  The following is a
generator-based algorithm which computes sup$\mathcal {C}(K)$
\cite{WonRam:87}.

\emph{Algorithm 2:} Input $\textbf{G}=(Q, \Sigma, \delta, q_0, Q_m)$
and $\textbf{K}=(Y,\Sigma,\eta,y_0,Y_m)$ representing $L_m({\bf G})$
and $K$, respectively.

\noindent 1. Set ${\bf K}_0 = (Y_0, \Sigma, \eta_0, y_0, Y_{m,0}) =
{\bf K}$.

\noindent 2. For $i \geq 0$, calculate ${\bf K}'_i = (Y'_i, \Sigma,
\eta'_i, y_0, Y'_{m,i})$ where
\begin{align*}
&Y'_i = \{ y \in Y_i \ |\ (\forall q \in Q)(\exists s \in L({\bf
K}_i))\ y=\eta(y_0,s), q=\eta(q_0,s), \Sigma(q) \cap \Sigma_u
\subseteq \Sigma(y) \}, \\
&\mbox{\hspace{0.8cm} where $\Sigma(\cdot)$ is the set of events defined at the argument state;} \\
&Y'_{m,i} = Y_{m,i} \cap Y'_i; \\
&\eta'_i = \eta_i |_{Y'_i}, \mbox{ the restriction of $\eta_i$ to
$Y'_i$.}
\end{align*}

\noindent 3. Set ${\bf K}_{i+1} =$ trim$({\bf K}'_i) = (Y_{i+1},
\Sigma, \eta_{i+1}, y_0, Y_{m,i+1})$.\footnote{If the initial state
$y_0$ has disappeared, the result is empty.} If ${\bf K}_{i+1} =
{\bf K}_{i}$, then output $\textbf{H} = {\bf K}_{i+1}$. Otherwise,
advance $i$ to $i+1$ and go to Step 2.

By \cite{WonRam:87} we know $L_m(\textbf{H}) = $ sup$\mathcal
{C}(K)$.  In each iteration of Algorithm~2, some states (at least
one) of ${\bf K}$, together with transitions incident on them, are
removed, either because the controllability condition is violated by
some string(s) reaching the states, or these states are
non-reachable or non-coreachable. Thus, the algorithm terminates in
at most $|Y|$ iterations.


Now we design an algorithm, which iteratively applies Algorithms~1
and 2, to compute a controllable and relatively observable
sublanguage of $K$. Let Assumptions~1 and 2 in
Section~\ref{sec_relobs-gen} hold.

\emph{Algorithm 3:} Input $\textbf{G}$, $\textbf{K}$, and
$P:\Sigma^* \rightarrow \Sigma^*_o$.

\noindent 1. Set ${\bf K}_0 = {\bf K}$.

\noindent 2. For $i \geq 0$, apply Algorithm~2 with inputs
$\textbf{G}$ and ${\bf K}_i$. Obtain ${\bf H}_i$ such that $L_m({\bf
H}_i) = $ sup$\mathcal {C}(L_m({\bf K}_i))$.

\noindent 3. Apply Algorithm~1 with inputs $\textbf{G}$, ${\bf
H}_i$, and $P:\Sigma^* \rightarrow \Sigma^*_o$. Obtain ${\bf
K}_{i+1}$ such that $L_m({\bf K}_{i+1}) =$ sup$\mathcal {O}(L_m({\bf
H}_i)) = $ sup$\mathcal {O}($sup$\mathcal {C}(L_m({\bf K}_i)))$. If
${\bf K}_{i+1} = {\bf K}_{i}$, then output $\textbf{K}_{\infty} =
{\bf K}_{i+1}$. Otherwise, advance $i$ to $i+1$ and go to Step~2.

Note that in applying Algorithm~1 at Step~3, the ambient language
successively shrinks to the supremal controllable sublanguage
sup$\mathcal {C}(L_m({\bf K}_i))$ computed by Algorithm~2 at the
immediately previous Step~2 of Algorithm~3. Thus every $L_m({\bf
K}_{i+1})$ is relatively observable with respect to sup$\mathcal
{C}(L_m({\bf K}_i))$. This choice of ambient languages is based on
the intuition that at each iteration $i$, any behavior outside
sup$\mathcal {C}(L_m({\bf K}_i))$ may be effectively disabled by
means of control, and hence is discarded when observability is
tested. The successive shrinking of ambient languages is useful in
computing less restrictive controlled behavior, as compared to the
algorithm in \cite{TakaiUshio:03} which is equivalent to fixing the
ambient language at $L({\bf G})$. An illustration is the Guideway
example in the next section.

Since Algorithms~1 and 2 both terminate in finite steps, and there
can be at most $|Y|$ applications of the two algorithms, Algorithm~3
also terminates in finite steps.  This means that the sequence of
languages
\begin{align*}
L_m({\bf K}_0) \supseteq L_m({\bf H}_1) \supseteq L_m({\bf K}_1)
\supseteq L_m({\bf H}_2) \supseteq L_m({\bf K}_2) \supseteq \cdots
\end{align*}
is finitely convergent to $L_m({\bf K}_\infty)$. The complexity of
Algorithm~3 is exponential in $|Y|$ because Algorithm~1 is of this
complexity.

Note that in testing the condition~(\ref{eq:c-obs1}) of relative
observability in Algorithm~3, we restrict attention only to
$\Sigma_c$ because uncontrollable transitions are dealt with by the
controllability requirement.

\begin{thm} \label{thm:sup-obs-con}
$L_m(\textbf{K}_{\infty})$ is controllable and observable, and
contains at least the supremal controllable and normal sublanguage
of $K$.
\end{thm}

\emph{Proof.} For the first statement, let ${\bf K}_{\infty} = {\bf
K}_{i+1} = {\bf K}_{i}$ for some $i \geq 0$.  According to Steps~2
and 3 of Algorithm~3, the latter equality implies that
$L_m(\textbf{K}_{\infty})$ is controllable and
$\overline{\mbox{sup}\mathcal {C}(L_m({\bf K}_{i}))}$-observable.
Therefore $L_m(\textbf{K}_{\infty})$ is controllable and observable
by Proposition~\ref{prop:c-obs_obs}.

To see the second statement, set up a similar algorithm to
Algorithm~3 but replace Step~3 by a known procedure to compute the
supremal normal sublanguage (\cite{ChoMarcus:89,BrandtLin:90}).
Denote the resulting generators by ${\bf K}'_i$.  Then by
Proposition~\ref{thm:sup-c-obs}, $L_m({\bf K}_i) = $ sup$\mathcal
{O}(\mbox{sup}\mathcal {C}(L_m({\bf K}_{i-1}))) \supseteq L_m({\bf
K}'_i)$, for all $i \geq 1$.  Now suppose the new algorithm
terminates at the $j$th iteration. Then Algorithm~3 must terminate
at the $j$th iteration or earlier, because normality implies
relative observability. Therefore $L_m({\bf K}'_j) \subseteq
L_m({\bf K}_j)$, i.e. $L_m(\textbf{K}_{\infty})$ contains the
supremal controllable and normal sublanguage of $K$. \hfill
$\square$

Algorithm~3 has been implemented as a procedure in \cite{TCT}. To
empirically demonstrate Theorem~\ref{thm:sup-obs-con}, the next
section applies Algorithm~3 to study two examples, Guideway and AGV.


%% file: sec_examples.tex
\section{Examples} \label{sec_examples}

Our first example, Guideway, illustrates that Algorithm~3 computes
an observable and controllable language larger either than the one
based on normality or that of \cite{TakaiUshio:03}. The second
example, the AGV system, provides computational results to
demonstrate Algorithm~3 as well as to compare relative observability
and normality.

\subsection{Control of a Guideway under partial observation} \label{sec_guideway}

\begin{figure}[!t]
  \centering
  \includegraphics[width=0.48\textwidth]{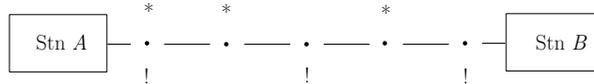}
  \caption{Guideway: stations A and B are connected by a single one-way track from A to B.
  The track consists of 4 sections, with stoplights ($*$) and detectors (!) installed at
  various section junctions as displayed.}
  \label{fig:guideway}
\end{figure}

\begin{figure}[!t]
  \centering
  \includegraphics[width=0.45\textwidth]{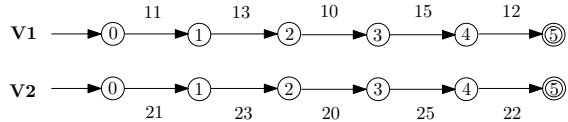}
  \caption{Vehicle generator model}
  \label{fig:vehicle}
\end{figure}

We demonstrate relative observability and Algorithm~3 with a
Guideway example, adapted from \cite[Section~6.6]{SCDES}. As
displayed in Fig.~\ref{fig:guideway}, stations A and B on a Guideway
are connected by a single one-way track from A to B. The track
consists of 4 sections, with stoplights ($*$) and detectors (!)
installed at various section junctions.  Two vehicles, ${\bf V}_1$
and ${\bf V}_2$, use the Guideway simultaneously.  Their generator
models are displayed in Fig.~\ref{fig:vehicle}; ${\bf V}_i$,
$i=1,2$, is at state $0$ (station A), state $j$ (while travelling in
section $j=1,...,4$), or state $5$ (station B).  The plant ${\bf G}$
to be controlled is ${\bf G} = {\bf V}_1 || {\bf V}_2$.

To prevent collision, control of the stoplights must ensure that
${\bf V}_1$ and ${\bf V}_2$ never travel on the same section of
track simultaneously: i.e. ensure mutual exclusion of the state
pairs $(j,j), j=1,...,4$. Let ${\bf K}$ be a generator enforcing
this specification.  Here according to the locations of stoplights
($*$) and detectors (!) displayed in Fig.~\ref{fig:guideway}, we
choose controllable events to be $i1,i3,i5$, and unobservable events
$i3,i5$, $i=1,2$. The latter define a natural projection $P$.

\begin{figure}[!t]
  \centering
  \includegraphics[width=0.58\textwidth]{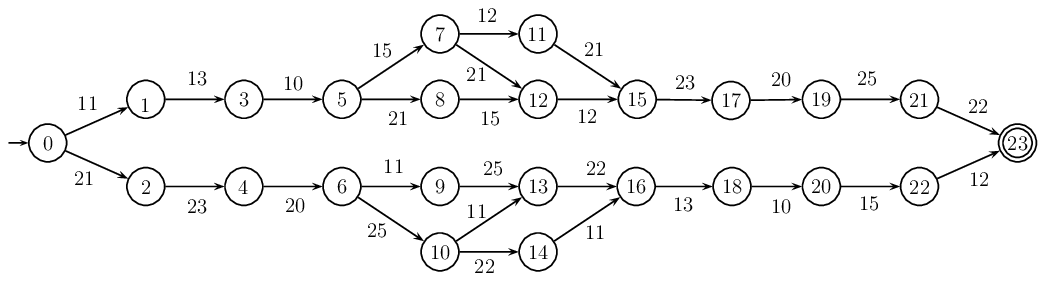}
  \caption{Supremal relatively observable and controllable sublanguage}
  \label{fig:supobs}
\end{figure}

\begin{figure}[!t]
  \centering
  \includegraphics[width=0.58\textwidth]{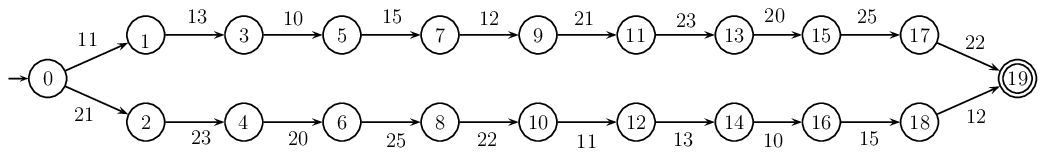}
  \caption{Supremal normal and controllable sublanguage}
  \label{fig:supnorm}
\end{figure}

First, applying Algorithm~2, with inputs ${\bf G}$, ${\bf K}$, and
$\Sigma_c$, we obtain the full-observation monolithic supervisor,
with 30 states, 40 transitions, and marked language sup$\mathcal
{C}(L_m({\bf G} || {\bf K}))$. Now applying Algorithm~3 we obtain
the generator displayed in Fig.~\ref{fig:supobs}; Algorithm~3
terminates after just one iteration. The resulting controlled
behavior is verified to be controllable and observable (as
Theorem~\ref{thm:sup-obs-con} asserts). Moreover, it is strictly
larger than the supremal normal and controllable sublanguage
represented by the generator displayed in Fig.~\ref{fig:supnorm}.
The reason is as follows. After string $11.13.10$, ${\bf V}_1$ is at
state $3$ (section $3$) and ${\bf V}_2$ at $0$ (station A).  With
relative observability, either ${\bf V}_1$ executes event $15$
(moving to state $4$) or ${\bf V}_2$ executes $21$ (moving to state
$1$); in the latter case, the controller disables event $23$ after
execution of $21$ to ensure mutual exclusion at $(3,3)$ because
event $20$ is uncontrollable. With normality, however, event $23$
cannot be disabled because it is unobservable; thus $21$ is disabled
after string $11.13.10$, and the only possibility is that ${\bf
V}_1$ executes $15$. In fact, $21$ is kept disabled until the
observable event $12$ occurs, i.e. ${\bf V}_1$ arrives at station~B.

For this example, the algorithm in \cite{TakaiUshio:03} yields the
same generator as the one in Fig.~\ref{fig:supnorm};  indeed, states
12 and 13 of the generator in Fig.~\ref{fig:supobs} must be removed
in order to meet the observability definition in
\cite{TakaiUshio:03}. Thus, this example illustrates that our
algorithm can obtain a larger controlled behavior compared to
\cite{TakaiUshio:03}.

%


\subsection{Control of an AGV System under partial observation} \label{sec_agv}

\begin{figure}[!t]
  \centering
  \includegraphics[width=0.28\textwidth]{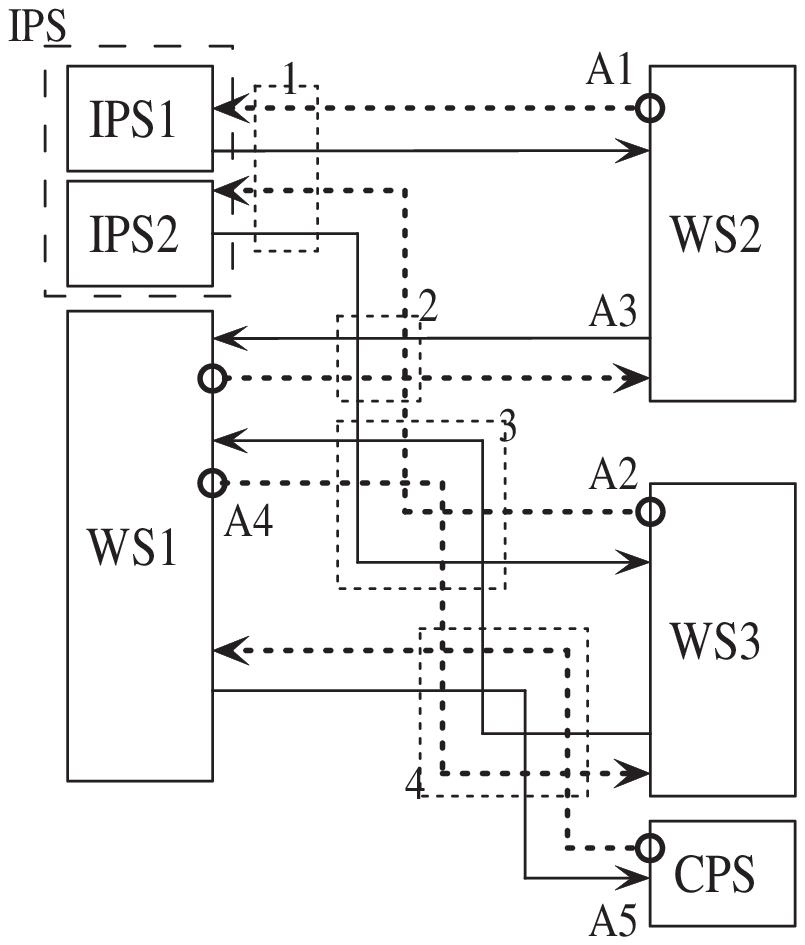}
  \caption{AGV: system configuration}
  \label{fig:agv}
\end{figure}
\begin{figure*}[!t]
  \centering
  \includegraphics[width=0.85\textwidth]{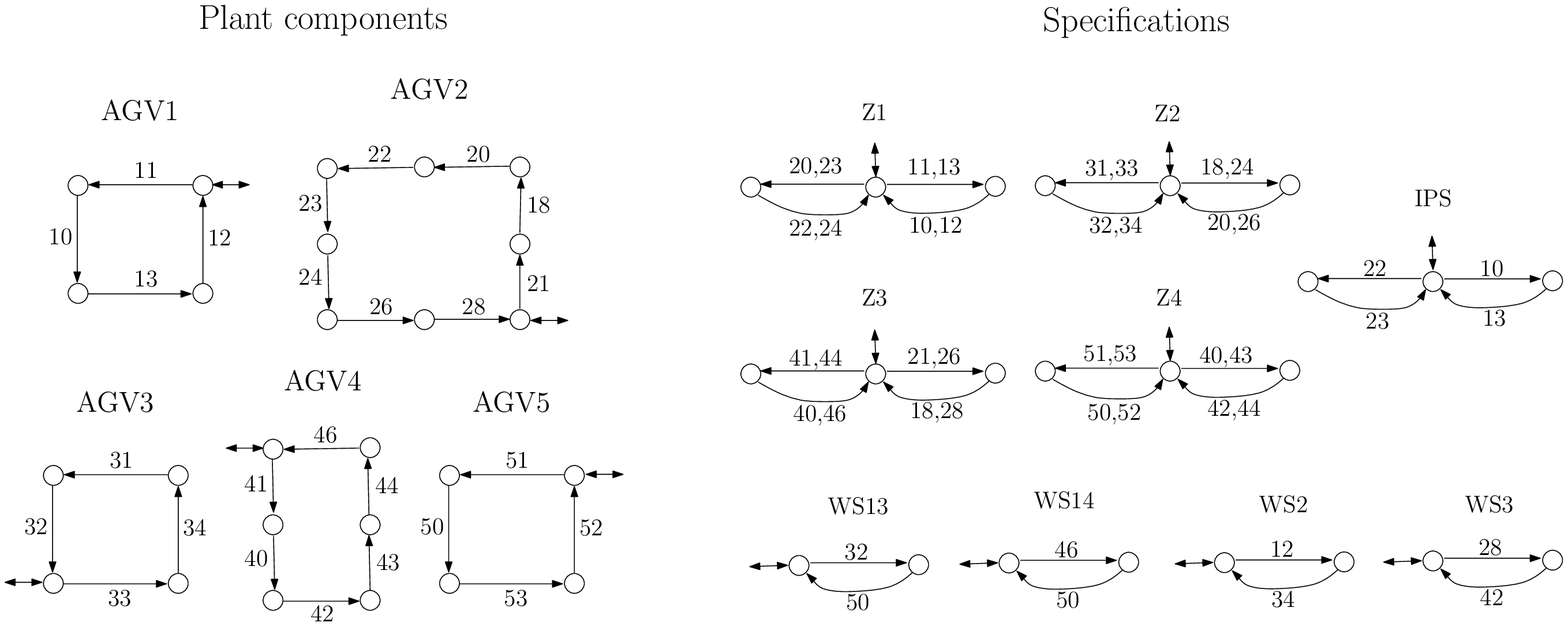}
  \caption{Generators of plant components and specifications}
  \label{fig:AGV_Plant_Spec}
\end{figure*}

We now apply Algorithm~3 to study a larger example, a system of five
automated guided vehicles (AGVs) serving a manufacturing workcell,
in the version of \cite[Section 4.7]{SCDES}, originally adapted from
\cite{HolKro:90}.

As displayed in Fig.~\ref{fig:agv}, the workcell consists of two
input parts stations IPS1, IPS2 for parts of types 1 and 2, three
workstations WS1, WS2, WS3, and one completed parts station CPS.
Five independent AGVs -- AGV1,...,AGV5 -- travel in fixed
criss-crossing routes, loading/unloading and transporting parts in
the cell. We model the synchronous product of the five AGVs as the
plant to be controlled, on which three types of control
specifications are imposed: the mutual exclusion (i.e., single
occupancy) of shared zones (dashed squares in Fig.~\ref{fig:agv}),
the capacity limit of workstations, and the mutual exclusion of the
shared loading area of the input stations. The generator models of
plant components and specifications are displayed in Fig.
\ref{fig:AGV_Plant_Spec}; here odd numbered events are controllable,
and there are 10 such events, $i1, i3$, $i=1,...,5$. For observable
events, we will consider different subsets of events below. The
reader is referred to \cite[Section 4.7]{SCDES} for the detailed
interpretation of events.

Under full observation, we obtain by Algorithm~2 the monolithic
supervisor of 4406 states and 11338 transitions. Then we select
different subsets of controllable events to be unobservable, and
apply Algorithm~3 to compute the corresponding supervisors which are
relatively observable and controllable. The computational results
are displayed in Table~\ref{tab:agv_relobs_norm}; the supervisors
are state minimal, and controllability, observability, and normality
are independently verified.  All computations and verifications are
done by procedures implemented in \cite{TCT}.

The cases in Table~\ref{tab:agv_relobs_norm} show considerable
differences in state size between relatively observable and
controllable supervisors and the normal counterparts. In the case
$\Sigma_{uo}=\{13\}$, the monolithic supervisor is in fact
observable in the standard sense; thus Algorithms~1 and 3 both
terminate after 1 iteration, and no transition removing or state
unmarking was done. By contrast, the normal supervisor loses 890
states. The contrast in state size is more significant in the case
$\Sigma_{uo}=\{21\}$: while the normal supervisor is empty, the
relatively observable supervisor loses merely 58 states compared to
the full-observation supervisor. The last row of
Table~\ref{tab:agv_relobs_norm} shows a case where only two out of
ten controllable events, 11 and 21, are observable. Still, relative
observability produces a 579-state supervisor, whereas the normal
supervisor is already empty when only events 41 and 51 are
unobservable (the third case). Finally, comparing the last two rows
of Table~\ref{tab:agv_relobs_norm} we see that making event 11
(``AGV1 enters zone1'') unobservable substantially reduces the
supervisor's state size, and, indeed, the effect is more substantial
than making six other events $\{13,23,33,43,51,53\}$ unobservable.
Such a comparison allows us to identify which event(s) may be
observationally critical with respect to controlled behavior.

Note from the state sizes of relatively observable supervisors in
Table~\ref{tab:agv_relobs_norm} that no state increase occurs
compared to the full-observation supervisor. In addition, the last
two columns of Table~\ref{tab:agv_relobs_norm} suggest that
Algorithm~3 with Algorithm~1 embedded terminates reasonably fast.

\begin{table}[!t]
\renewcommand{\arraystretch}{1.3}
\caption{Test results of Algorithm~3 for different subsets of
unobservable events in the AGV system} \label{tab:agv_relobs_norm}
\centering
\begin{tabular}{|c||c|c|c|c|}\hline
$\Sigma_{uo}=\Sigma-\Sigma_o$ & State \# of rel. obs. supervisor  & State \# of normal supervisor & Iteration \# of Alg.~3 & Iteration \# of Alg.~1 \\
\hline
\{13\} & 4406 & 3516 & 1 & 1 \\
\hline
\{21\} & 4348 & 0 & 1 & 399 \\
\hline
\{41,51\} & 3854 & 0 & 2 & 257 \\
\hline
\{31,43\} & 4215 & 1485 & 1 & 233 \\
\hline
\{11,31,41\} & 163 & 0 & 1 & 28 \\
\hline
\{13,23,31,33, & \multirow{2}{*}{579} & \multirow{2}{*}{0} & \multirow{2}{*}{3} & \multirow{2}{*}{462} \\
41,43,51,53\} & & & & \\
\hline
\end{tabular}
\end{table}


%% file: sec_concl.tex
\section{Conclusions} \label{sec_concl}

We have identified the new concept of relative observability, and
proved that it is stronger than observability, weaker than
normality, and preserved under set union. Hence there exists the
supremal relatively observable sublanguage of a given language. In
addition we have provided an algorithm to effectively compute the
supremal sublanguage.

Combined with controllability, relative observability generates
generally larger controlled behavior than the normality counterpart.
This has been demonstrated with a Guideway example and an AGV
example. Empirical results for the AGV example show considerable
improvement of controlled behavior using relative observability as
compared to normality.

Newly identified, the algebraically well-behaved concept of relative
observability may be expected to impact several closely related
topics such as coobservability, decentralized supervisory control,
stated-based observability, and observability of timed
discrete-event systems. In future work we aim to explore these
directions.